\documentclass[axioms,article,accept,pdftex,moreauthors]{Definitions/mdpi} 

\def\G{\Gamma}
\def\no{\nonumber}
\def\rar{\rightarrow}

\def\le{\left(}
\def\ri{\right)}

\def\res{\mathop{{\rm Res}}\limits}
\firstpage{1} 
\pubvolume{1}
\issuenum{1}
\articlenumber{0}
\pubyear{2025}
\copyrightyear{2024}
\datereceived{ } 
\daterevised{ } 
\dateaccepted{ } 
\datepublished{ } 
\hreflink{https://doi.org/} 

\Title{A Simple Way to Reduce the Number of Contours in the Multi-Fold Mellin-Barnes Integrals}

\TitleCitation{A Simple Way to Reduce the Number of Contours in the Multi-Fold Mellin-Barnes Integrals}

\Author{Mauricio Diaz $^{1}$, Ivan Gonzalez $^{2,3}$, Igor Kondrashuk $^{3,4}$ and Eduardo A. Notte-Cuello $^{3,5}$}

\AuthorNames{Mauricio Diaz, Ivan Gonzalez, Igor Kondrashuk and Eduardo A. Notte-Cuello}

\isAPAStyle{%
       \AuthorCitation{Lastname, F., Lastname, F., \& Lastname, F.}
         }{%
        \isChicagoStyle{%
        \AuthorCitation{Lastname, Firstname, Firstname Lastname, and Firstname Lastname.}
        }{
        \AuthorCitation{Diaz, M.; Gonzalez, I.; Kondrashuk, I.; Notte-Cuello, E.A.}
        }
}

\address{%
$^{1}$ \quad Departamento de F\'isica, Universidad del B\'io-B\'io, Av. Collao 1202,  Casilla 15-C,
     Concepci\'on, Chile\\
$^{2}$ \quad Instituto de F\'isica y Astronom\'ia, Universidad de Valpara\'iso, Av. Gran Breta\~na
       1111,  Valpara\'iso, Chile\\
$^{3}$ \quad Centro de Ciencias Exactas, Universidad del B\'io-B\'io,  Campus Fernando May, 
       Av. Andres Bello 720, Casilla 447, Chill\'an, Chile\\
$^{4}$ \quad Grupo de Matem\'atica Aplicada {\rm \&} Grupo de F\'isica de Altas Energ\'ias, 
      Departamento de Ciencias B\'asicas, Universidad del B\'io-B\'io,  Campus Fernando May, 
      Av. Andres Bello 720, Casilla 447, Chill\'an, Chile\\
$^{5}$ \quad Departamento de Matem\'aticas, Facultad de Ciencias, Universidad de La Serena, 
          Av. Cisternas 1200, La Serena, Chile}

\abstract{Mellin-Barnes integral representation  of one-loop off-shell box massless diagram  is five-fold by construction. On the other hand, 
it is known from the year 1992 that it may be reduced to certain two-fold Mellin-Barnes integral.  We  propose a way to reduce the number of the Mellin-Barnes integration contours from five to two by using the Mellin-Barnes integral representation only in combination with    basic methods of mathematical analysis such as analytical regularization.  We do not use any Barnes lemma  to prove the reduction but we use the integral Cauchy formula instead.  We recover first the well-known two-fold Mellin-Barnes representation for the one-loop triangle massless diagram and then show how the five-fold Mellin-Barnes integral representation of one-loop box diagram with all the indices 1 in four spacetime dimensions may be reduced to the two-fold Mellin-Barnes representation for one-loop triangle diagram.  Singular integrals over Feynman parameters appear in the integrand of the five-fold Mellin-Barnes integral representation at the intermediate step. Such integrals should be treated as
distributions with respect to certain linear combinations of the initial Mellin-Barnes integration variables  in the Mellin-Barnes integrands.  These distributions may be integrated out with a finite number of residues in the limit of removing the analytical regularization.  We explain how to apply this strategy to an arbitrary Feynman diagram in order to reduce the number of Mellin-Barnes integration contours. On the practical  side,  we analyze connections between the obtained results and the knot theory,  Trotter integrals, quantum computing.}


\begin{document}

\section{Introduction}

Significant progress has been made in theoretical particle physics  with purpose to study scattering amplitudes. Major progress occurred after the BDS conjecture  \cite{Bern:2005iz,Bern:2006ew} for the amplitudes of four particle processes in maximally symmetric $ {\cal N}=4$  SYM theory. However, comparatively fewer studies have focused on Green functions, the quantities obtained from the path integral when expressed in terms of external sources.  The first results were obtained by Usyukina and Davydychev \cite{ Usyukina:1992jd,  Usyukina:1993ch} in the context of massless ladder diagrams, where they found expressions in terms of polylogarithms and demonstrated the equivalence of three-point and four-point Green functions using a combination of Feynman parameters, Mellin-Barnes integral representations, and the uniqueness trick.   The equivalence between four-point and three-point ladder diagrams may also be demonstrated via the Jacobian of a conformal transformation in an auxiliary space dual to the four-dimensional momentum space  \cite{Kondrashuk:2008ec,Kondrashuk:2008xq,Kondrashuk:2009us,Broadhurst:1993ib}.  Usyukina and Davydychev's results were based on the loop reduction technique proposed by Belokurov and Usyukina in 1983 for the four-dimensional case  \cite{Belokurov:1983km}. This loop reduction technique was generalized for non-integer dimensions in \cite{Gonzalez:2012gu} and \cite{Gonzalez:2012wk}  by two of us, although this was specifically the case for conformal ladders, which is not a practical case because the index had to be modified to make the Belokurov-Usyukina loop reduction technique applicable when the dimension of spacetime is arbitrary.

The three-point  and four-point ladders contribute to the Green functions of rank three and four, respectively. The Green functions can  be obtained from the path integral, the logarithm of which is related to the effective action by a Legendre transformation \cite{Kondrashuk:2000qb}.  The path integral and the effective action are restricted by the Slavnov-Taylor identity \cite{Kondrashuk:2000qb}. In \cite{Kondrashuk:2004pu},  it was shown that the effective action of the dressed mean fields can be treated as the classical action but with the indices of the dressing function depending on two complex variables.
These two complex indices are integrated with a  function of two complex variables, which determines the auxiliary double ghost vertex in the Landau gauge. This complex function of two  variables can  be determined by  Bethe-Salpeter equation \cite{Allendes:2009bd,Kondrashuk:2004pu}.  There is a  reason to expect an  iterative structure for this complex function of two variables that solves the BS equation.

The reason for the  simple iterative structure of this double ghost vertex is that the structure of the three-gluon vertex  in  $ {\cal N}=4$  SYM theory  is fixed by conformal invariance in the Landau gauge \cite{Cvetic:2004kx,Cvetic:2006kk,Cvetic:2006iu,Cvetic:2007ds,Cvetic:2007fp}.     This simple structure is mapped onto the structure of the $Lcc$ vertex by the ST identity,  which cannot be fixed by conformal invariance because the auxiliary field $L$ does not appear in the integration measure of the path integral \cite{Cvetic:2006iu}.   Despite  this,  the structure of the double ghost vertex is expected to remain simple at all loops because the the structure of the three gluon vertex is simple in any dimension,  assuming the well-known arguments based on anomalies, even in arbitrary dimensions \cite{Kondrashuk:2004pu}. For QED, the restriction imposed by conformal invariance in  position space was considered in \cite{Fradkin:1978pp,Palchik:1982ta,Mitra:2008pw,Mitra:2008yr,Mitra:2009zm}.  Conformal symmetry fixes the three-gluon vertex of the dressed gluons in the Landau gauge in the integer dimensions \cite{Cvetic:2007ds,Kondrashuk:2004pu}.
This suggests that the structure of the double ghost vertex should also be simplified in  any  number of dimensions  \cite{Allendes:2009bd,Allendes:2012mr,Kondrashuk:2004pu}.

 On the other hand, the BS  equation for this auxiliary vertex has a relatively simple structure because the ghosts interact only with the gauge fields \cite{Cvetic:2002dx,Cvetic:2002in,Allendes:2009bd,Allendes:2012mr}.  The $Lcc$ vertex was calculated in the Landau gauge because it does not have any superficial divergence in this gauge. It was calculated at the two-loop level in the $d=4$ case; however, in four dimensions, it is impossible to put the Green functions on-shell to compare with the known results for scattering amplitudes \cite{Kondrashuk:2004pu}.  We need a regularization procedure to put the results of the calculation for the correlators on-shell and to regularize the amplitudes in the infrared limit. In \cite{Allendes:2009bd,Gonzalez:2018gch,Kondrashuk:2004pu}, a very simple representation for this vertex was found, which is valid in any number of dimensions and for any gauge. 
  Specifically, this vertex can be represented as a star-like integral, where the indices of the star rays depend on two complex variables. Starting from this structure, we can recover the remaining Green functions according to the approach established in 
 \cite{Cvetic:2002dx,Cvetic:2002in,Kondrashuk:2000br}.

The simplicity of the structure of the $Lcc$ vertex suggests that a consistent reduction of Mellin-Barnes integrals can be searched in the Mellin-Barnes representation for subgraphs of this vertex. Typically, such reductions in Mellin-Barnes integral representations of Feynman diagrams arise from Barnes lemmas or their derivatives \cite{Smirnov,Barnes-1,Barnes-2}, which imply loop reduction \cite{Allendes:2012mr,Kniehl:2013dma,Gonzalez:2016pgx}. However, there are cases where Mellin-Barnes integrals can be reduced without depending on the Barnes lemmas.  For example, the reduction of the rank of the four-point ladder contribution to the Green function of rank three does not involve the Barnes lemma but rather relies on basic methods of mathematical analysis. This will be demonstrated in the present paper.  The reason for such a reduction in Mellin-Barnes integrals will be explained. The method produces unexpected relations between different hypergeometric functions. Although this is not a new concept since the rank reduction of Green functions has been known for over 30 years, our analysis allows for the generation of more explicit relations. We will show in this paper why this occurs and how it can be used to reduce the number of Mellin-Barnes contours in more general cases.  The key reason is that several specific values of linear combinations of the original complex variables in the Mellin-Barnes transform contribute to the Mellin-Barnes integrals corresponding to Feynman diagrams. The remaining residues for such combinations do not contribute. This conclusion arises from analyzing the integration over Feynman parameters. It resembles the method of brackets \cite{Gonzalez:2021vqh}.

As a result of this strategy, in multi-fold Mellin-Barnes calculations, we first reduce the rank of the contributions to the Green functions using our trick derived from basic methods of mathematical analysis. We then apply the Barnes lemmas to reduce a multi-fold Mellin-Barnes integral to a two-fold Mellin-Barnes integral \cite{Kniehl:2013dma}.  Typically, this reduction in the number of Mellin-Barnes contours is accomplished by combining complex integration with standard Riemann integration \cite{Alvarez:2019eaa}. However, in this work, we adopt a different strategy and carry out the reduction using Mellin-Barnes integrals alone.

In general, the number of contours in the Mellin-Barnes integral representation of a Feynman diagram is first related to the rank of the Green function to which the diagram contributes, and second, to the number of loops in the diagram. The rank of the Green function corresponds to the number of momenta that enter the diagram. A four-leg diagram, for example, may effectively be reduced to a three-leg diagram. We now raise the question: Can we reduce the number of Mellin-Barnes integrations in a multi-leg Mellin-Barnes integral using complex integration alone, without relying on intermediate Riemann integrals as in \cite{Alvarez:2019eaa} or without performing any diagram transformation in the dual space from   \cite{Kondrashuk:2009us}?
 The fact that previous papers sometimes combined Riemannian integration and Mellin-Barnes integration suggests that more efficient formulas in terms of complex variables alone have yet to be found.

The paper is organized as follows.    Section \ref{sec-MMellin-BarnesT}  provides a brief review of the Mellin-Barnes transformation.  It is the main tool of the present research. The special attention is given to the multi-fold Mellin-Barnes transformation as the primary goal of the paper is the number of the Mellin-Barnes integration contours in the integral representation of the box diagram.    In Section  \ref{sec-FP-triangle} we consider the
Feynman formula derived in terms of the Euler beta function and apply it to the triangle massless diagram  in reproducing  the uniqueness case.   In Section  \ref{sec-out-Mellin-Barnes}  that is the main section of the present paper the Mellin-Barnes transformation is applied to the denominator of the integral over the  Feynman parameters of two Feynman diagrams.  First, in Subsection \ref{subsec-two-fold-triangle}   we apply it to the integral over the Feynman parameters obtained for the triangle massless diagram with arbitrary indices in arbitrary spacetime dimension and recover the well-known two-fold Mellin-Barnes representation for this triangle diagram.   The convergence domain for the indices of propagators and dimension of the spacetime is established.  Second, in Subsection \ref{subsec-five}   we apply the Mellin-Barnes transformation to the denominator of the integrand over the Feynman parameters obtained for the massless box diagram in $d=4$ and all the indices equal to 1 and come to a five-fold Mellin-Barnes integral.  In Subsection \ref{subsec-five}  we show how the five-fold Mellin-Barnes integral can be reduced to the two-fold Mellin-Barnes integral by the basic methods of mathematical analysis with help of an analytical regularization.  In Section \ref{sec-Discuss} we generalize this strategy learned from the massless box diagram to an arbitrary Feynman diagram.  Appendix \ref{App} repeats the way of Usyukina and Davydychev \cite{Usyukina:1992jd} from the box  back to the triangle diagram.  It is necessary to compare with our method for which only the multi-fold  Mellin-Barnes representation is necessary.

\section{Multi-fold Mellin-Barnes transformation} \label{sec-MMellin-BarnesT}

A brief review and a simple summary  \footnote{
All this Section \ref{sec-MMellin-BarnesT} of the present article is based on the lectures titled “Integración multiple y sus aplicaciones”
given by I.K. at the Mathematical Department of Facultad de Educación y Humanidades, UBB, Chillán, during the second semesters of 2010, 2018, 2020, 2022, and 2023} of the  Mellin transformation with respect to one variable
may be found in \cite{Gonzalez:2021vqh,Alvarez:2016juq}.  One-fold Mellin-Barnes transformation is a particular case of the Mellin
transformation. Also, in \cite{Gonzalez:2021vqh,Alvarez:2016juq} the asymptotic behaviors of Mellin moment and Mellin transform were compared.
The main purpose of the present paper is to  reduce multi-fold  Mellin-Barnes integrals to  two-fold Mellin-Barnes integrals.
We need to recall in this Section in brief the main steps of Mellin transformation in order to generalize one-fold Mellin-Barnes transformation to a multi-fold Mellin-Barnes transformation because we use extensively the multi-fold Mellin-Barnes transformation in this paper. Indeed, the Mellin transformation  is an integral transformation which is defined as
\begin{eqnarray*} 
M[f(x),x](z) = \int\limits_0^{\infty} x^{z-1}f(x)~dx,
\end{eqnarray*}
in which the arguments in the brackets on the l.h.s. stand for the transforming function $f(x)$  and the integration variable $x$ of
this integral transformation. The inverse Mellin transformation is
\begin{eqnarray} \label{MT-back}
f(x) = \frac{1}{2\pi i}\int\limits_{c-i\infty}^{c + i\infty}x^{-z} M[f(x),x](z) ~dz.
\end{eqnarray}
The position point $c$ of the  vertical line  of the integration contour
in the complex plane must be in the vertical strip $c_1 < c < c_2,$ the borders of the strip are defined by the condition that two integrals
\begin{eqnarray*}
\int\limits_0^{1} x^{c_1-1}f(x)~dx  &  {\rm and } &  \int\limits_1^{\infty} x^{c_2-1}f(x)~dx
\end{eqnarray*}
must be finite. Should the contour in  Eq.(\ref{MT-back}) be closed to the left complex infinity or to the right complex infinity
depends on the explicit asymptotic behavior of the Mellin transform  $M[f(x),x](z)$ at the complex infinity.
We close to the left if the left complex infinity does not contribute and we close to the right if the right complex infinity does not contribute. Under this condition the original function $f(x)$ may be reproduced  via calculation of the residues by Cauchy formula. The simplest example of the Mellin transformation would be
\begin{eqnarray*}
\G(z) = \int\limits_0^{\infty} e^{-x}x^{z-1}~dx & {\rm and }  &   e^{-x} = \frac{1}{2\pi i}\int\limits_{c-i\infty}^{c + i\infty}x^{-z} \G(z) ~dz.
\end{eqnarray*}
The contour in the complex plane is the vertical line with ${\rm Re}~z = c$  in the strip $0 < c < A,$ where $A$ is a real and positive number, the contour must be closed to the left infinity, otherwise the right positive infinity would contribute due to the asymptotic behavior of $\G$ function.

In comparison, in the Mellin-Barnes transformation which we consider in the next paragraphs we choose to which infinity the contour should be closed by taking into account the absolute value of $x$ in  (\ref{MT-back}) because the Mellin-Barnes transform has already an established structure in terms  of the Euler $\G$ functions.  Indeed, the Mellin-Barnes transformation is the Mellin transformation of the function $f(x)=1/{(1+x)^{\lambda}},$ $\lambda$ is a complex number,
\begin{eqnarray} \label{Mellin-BarnesT}
\int\limits_0^{\infty} \frac{x^{z-1}}{(1+x)^{\lambda} } ~dx =   \int\limits_0^{\infty} \frac{x^{z-1}}{(1+x)^{z + (\lambda -z)}} ~dx = {\rm B}(z,\lambda-z) = \frac{\G(z)\G(\lambda -z) }{\G(\lambda)}.
\end{eqnarray}
From the construction of this integral (\ref{Mellin-BarnesT}) we obtain the condition $0 < {\rm Re}~z <    {\rm Re}~ \lambda.$ In this strip the integral (\ref{Mellin-BarnesT}) is convergent. The inverse transformation is
\begin{eqnarray} \label{Mellin-BarnesT-back}
\frac{1}{(1+x)^{\lambda}} = \frac{1}{2i\pi}  \frac{1}{\G(\lambda)} \int\limits_{c-i\infty}^{c + i\infty}x^{-z} \G(z)\G(\lambda -z)~dz.
\end{eqnarray}
Here $0 < c <  {\rm Re}~ \lambda,$ usually taken a bit to the right from the point $z=0.$  However, the more traditional presentation of the inverse transformation  (\ref{Mellin-BarnesT-back}) is to do the reflection in the plane of the complex variable $z \rar -z$ in the integrand of Eq. (\ref{Mellin-BarnesT-back}),  the inverse Mellin-Barnes transformation takes a form
\begin{eqnarray} \label{Mellin-BarnesT-back-2}
\frac{1}{(1+x)^{\lambda}} = \frac{1}{2\pi i}  \frac{1}{\G(\lambda)} \int\limits_{-c-i\infty}^{-c + i\infty}x^{z} \G(-z)\G(\lambda +z) ~dz,
\end{eqnarray}
this is more traditional form of the inverse Mellin-Barnes transformation and the straight vertical line passes a bit to the left from  $z=0$ because of   $0 > -c > - {\rm Re}~ \lambda.$ The inverse transformations   (\ref{Mellin-BarnesT-back}) or  (\ref{Mellin-BarnesT-back-2})  may be continued analytically to all the complex values of $\lambda$ and it becomes  valid even for $\lambda = -n,$ $n \in \mathbb{N}.$ In the latter case instead of the infinite series of residues we obtain a finite sum of them and reproduce the formula for the binomial. In any case we may write instead of $c$ in the Mellin-Barnes inverse transformation  (\ref{Mellin-BarnesT-back-2})  a small real positive $\delta$ for any complex value of $\lambda.$ If ${\rm Re}~ \lambda < 0$ the contour may be curved a bit in order to separate the
poles produced by $\G(\lambda +z)$ from the poles produced by the $\G(-z),$ that is, in order to separate ``left'' and ``right'' poles. Thus,  the contour passes between the leftmost pole $z=0$ of the  right poles of the integrand which are produced by $\G(z)$ and the rightmost pole $z=-\lambda$  of the left poles which are produced by  $\G(z+\lambda).$

The Mellin-Barnes transformation considered in the previous paragraphs can be generalized for any number of terms in the denominator.  The number  of the Mellin-Barnes integrations in such a case is the number of terms in the denominator  minus one. For three terms in the denominator we obtain  a two-fold Mellin-Barnes integrals. Consequent application of the Mellin-Barnes transformations creates the hierarchy of the contour positions. All the results done in this paper are based on this hierarchy of the positions of the contours.
Traditionally,  multi-fold Mellin-Barnes transformations generalize
the one-fold Mellin-Barnes transformation in the form (\ref{Mellin-BarnesT-back-2}) but not in the form (\ref{Mellin-BarnesT-back}) that is with the positive sign of the complex powers in the Mellin-Barnes integrands. \footnote{We omit the factor $1/2\pi i$ in front of each contour integral in the complex plane. This factor is always canceled  by the factor $2i\pi $ which appears in front of each residue due to Cauchy integral formula.}
 For example, for three terms in the denominator, we can write

\begin{eqnarray}  \label{two-fold-1}
\frac{1}{(A+B+C)^{\lambda}} =   \frac{1}{\G(\lambda)}   \int\limits_{-\delta_1-i\infty}^{-\delta_1 + i\infty} \G(-z_1)\G(\lambda +z_1)  \frac{B^{z_1} }{\le A+C\ri^{z_1+\lambda}}  ~dz_1 \no\\
= \frac{1}{\G(\lambda)}   \frac{1}{A^\lambda}  \int\limits_{-\delta_1-i\infty}^{-\delta_1 + i\infty} \G(-z_1)\G(\lambda +z_1) \le\frac{B}{A}\ri^{z_1} ~dz_1 \int\limits_{-\delta_2-i\infty}^{-\delta_2 + i\infty} \frac{\G(-z_2)\G(\lambda+z_1 +z_2)}{\G(z_1 + \lambda)}\le\frac{C}{A}\ri^{z_2}~dz_2  \no\\
= \frac{1}{\G(\lambda)}   \frac{1}{A^\lambda}  \int\limits_{-\delta_1-i\infty}^{-\delta_1+ i\infty}~dz_1 \int\limits_{-\delta_2-i\infty}^{-\delta_2 + i\infty}~dz_2\G(-z_2)\G(-z_1)
\G(\lambda +z_1+z_2) \le\frac{B}{A}\ri^{z_1}\le\frac{C}{A}\ri^{z_2}
\end{eqnarray}
Here the integration over the $z_2$ variable is done taking into account that ${\rm Re}~ z_1 =-\delta_1.$ According to the construction considered in (\ref{Mellin-BarnesT-back}) and (\ref{Mellin-BarnesT-back-2}) we should write  ${\rm Re}~\lambda-\delta_1 > \delta_2 > 0.$

\section{Feynman parameters for the triangle diagram} \label{sec-FP-triangle}

\begin{figure}[H] 
\centering\includegraphics[scale=0.5]{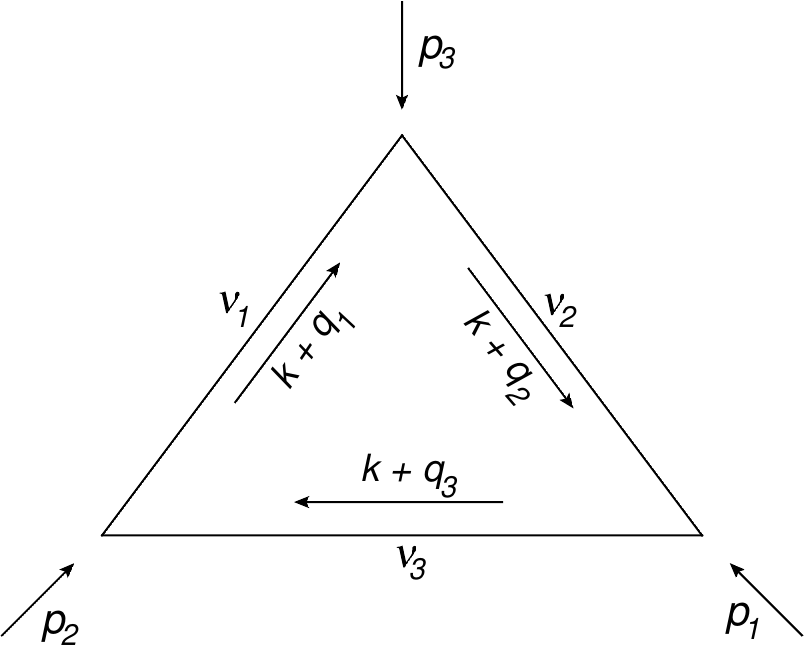}
\caption{\footnotesize  One-loop massless scalar triangle in momentum space}
\label{figure-1}
\end{figure}

Here we write a definition for the triangle scalar massless diagram with arbitrary indices \footnote{
All this Section \ref{sec-FP-triangle} is based on the QFT lectures given by  I.K. at UdeC, Chile from December of 2007 till November of 2012.}.  A one-loop massless triangle diagram is depicted in Fig. \ref{figure-1}.  It contains three scalar propagators. The $d$-dimensional momenta $p_1$, $p_2$, $p_3$ enter this diagram.
They are related by  momentum conservation
\begin{eqnarray*} 
p_1 + p_2 + p_3 = 0.  
\end{eqnarray*}
This momentum integral 
\begin{eqnarray} \label{triangle-UD}
J(\nu_1,\nu_2,\nu_3) = \int~Dk~\frac{1}{\left[(k + q_1)^2\right]^{\nu_1} \left[(k + q_2)^2 \right]^{\nu_2}
\left[(k + q_3)^2\right]^{\nu_3}}
\end{eqnarray}
corresponds to the diagram in Fig. \ref{figure-1}. The running momentum $k$ is the integration variable. The notation is chosen in such a way that the index of propagator $\nu_1$ stands on the line opposite to the vertex of triangle into which the momentum $p_1$ enters. This is a traditional definition of the triangle massless momentum integral \cite{Usyukina:1992jd,Usyukina:1993ch,Allendes:2012mr}.

The notation $q_1,$ $q_2$ and $q_3$ are taken from Ref.\cite{Usyukina:1992jd}.
It follows from the diagram in Fig. \ref{figure-1} and the momentum conservation law that  
\begin{eqnarray*}
p_1 = q_3 - q_2, ~~~ 
p_2 = q_1 - q_3, ~~~
p_3 = q_2 - q_1. 
\end{eqnarray*}
To define the integral measure in momentum space, we use the notation from Ref. \cite{Cvetic:2006iu}
\begin{eqnarray*}
Dk \equiv \pi^{-\frac{d}{2}}d^d k.    
\end{eqnarray*} 
Such a definition of the integration measure in momentum space helps to avoid powers of $\pi$ in formulas for the momentum integrals which will appear in the next formulas.

\subsection{Set of integrals} \label{subsec-Set}

Here we put a set of integrals which are behind the trick with Feynman parameters. All these integrals are three-hundred year old and were introduced in mathematics by Euler.  We need them because we will use them in the rest of this paper.
\begin{eqnarray} \label{beta}
\int\limits_{0}^{1} t^{u-1}(1-t)^{w-1} dt = \frac{\G(u)\G(w)}{\G(u+w)}, ~~~{\rm Re}~u >0, {\rm Re}~w >0
\end{eqnarray}
This integral (\ref{beta}) is convergent at both the limits for positive real parts of $u$ and $w.$ This simple old  integral which defines the Euler beta function will be the main element in the constructions used in all this paper. By changing the integration variable
\begin{eqnarray*} \label{beta+sub}
t = \frac{\tau}{1+\tau}
\end{eqnarray*}
we obtain for ${\rm Re}~u >0, {\rm Re}~w >0$
\begin{eqnarray} \label{beta+sub+gamma}
\int\limits_{0}^{1} t^{u-1}(1-t)^{w-1} dt =   \int\limits_{0}^{\infty} \frac{\tau^{u-1}}{(1+\tau)^{u+w}} d\tau =
\frac{\G(u)\G(w)}{\G(u+w)}.
\end{eqnarray}
This second integral in Eq. (\ref{beta+sub+gamma}) is more useful in order to be modified  because its limits of integration remain unchanged  with respect to  multiplications,
\begin{eqnarray} \label{multiplic}
\int\limits_{0}^{\infty} \frac{\tau^{u-1}}{(A+B\tau)^{u+w}} d\tau = \frac{1}{A^{u+w}} \int\limits_{0}^{\infty}
\frac{\tau^{u-1}}{(1+B\tau/A)^{u+w}} d\tau \no\\
= \frac{1}{A^w B^u} \int\limits_{0}^{\infty} \frac{\tau^{u-1}}{(1+\tau)^{u+w}} d\tau = \frac{1}{A^w B^u} {\rm B}(u,w).
\end{eqnarray}
We  obtain  from (\ref{multiplic}) that
\begin{eqnarray} \label{beta-2}
\int\limits_{0}^{1} \frac{t^{u-1}(1-t)^{w-1} }{(A+Bt)^{u+w}} dt =  \int\limits_{0}^{\infty}
\frac{\tau^{u-1}}{(1+\tau)^{u+w}} \frac{1}{\left[A + B\tau/(1+\tau)\right]^{u+w}}   d\tau \no\\
= \int\limits_{0}^{\infty} \frac{\tau^{u-1}}{\left[A(1+\tau) + B\tau\right]^{u+w}} d\tau = \frac{1}{A^w (A+B)^u} {\rm B}(u,w).
\end{eqnarray}
As a consequence of (\ref{beta-2}), we can write the relation
\begin{eqnarray} \label{beta-3}
\int\limits_{0}^{1} \frac{t^{u-1}(1-t)^{w-1} }{\left[At+B(1-t)\right]^{u+w}} dt
=  \frac{1}{B^w A^u} {\rm B}(u,w).
\end{eqnarray}
or equivalently, in the final form the integral (\ref{beta-3})
\begin{eqnarray} \label{feynman-2}
\frac{1}{B^w A^u} =
\frac{\G(u+w)}{\G(u)\G(w)}\int\limits_{0}^{1} \int\limits_{0}^{1} \frac{\alpha_1^{u-1}\alpha_2^{w-1}\delta(\alpha_1+\alpha_2-1) }{\left[A\alpha_1+B\alpha_2\right]^{u+w}} d\alpha_1d\alpha_2.
\end{eqnarray}
In Eq. (\ref{feynman-2}) for $\alpha_1$ and $\alpha_2$ the term ``Feynman parameters'' is used. The formula (\ref{feynman-2})
can be iteratively extended by induction to an arbitrary number of factors in the denominator on the left hand side. The right hand sides of Eqs. (\ref{beta}), (\ref{multiplic}), (\ref{beta-2}), (\ref{beta-3})
can be analytically continued. We cannot say the same about the left hand sides of these equations.  Some modified
integrals  should appear on the left hand sides when we continue analytically the right hand sides. For the values of the propagator indices $\nu_i$ that we use in this paper we are in the domain in which the traditional representation of the Feynman formula (\ref{feynman-2}) is
valid.

\subsection{Triangle integral over Feynman parameters}

We take representation of the integral (\ref{triangle-UD}) in terms of Feynman parameters  from Usyukina and Davydychev paper \cite{Usyukina:1992jd},
\begin{eqnarray} \label{triangle-FP}
J(\nu_1,\nu_2,\nu_3) = \frac{\G\le\sum\limits_{i=1}^3 \nu_i - \frac{d}{2}\ri}{\prod\limits_{i=1}^3 \G(\nu_i) }\int\limits_0^1\int\limits_0^1\int\limits_0^1\frac{\delta\le\sum\limits_{i=1}^3 \alpha_i -1\ri\prod\limits_{i=1}^3\alpha_i^{\nu_i-1}d\alpha_i}{(\alpha_1\alpha_2p_3^2 
+ \alpha_2\alpha_3 p_1^2 + \alpha_1\alpha_3 p_2^2 )^{ \sum\limits_i \nu_i - d/2}}
\end{eqnarray}
It can be obtained by applying the Feynman formula for three factors in the denominator to the triangle integral (\ref{triangle-UD}).

\subsection{Uniqueness case in terms of Feynman parameters} \label{subsec-uniq}

The uniqueness case $ \sum\limits_{i=1}^3 \nu_i =d$ for the triangle integral can be evaluated  using the
formulas of Subsection \ref{subsec-Set}. We obtain

\begin{adjustwidth}{-\extralength}{0cm}
\centering 
\begin{eqnarray*}
J(\nu_1,\nu_2,\nu_3) = \frac{\G(d/2)}{\prod\limits_i \G(\nu_i) }\int\limits_0^1\int\limits_0^1\int\limits_0^1\frac
{\delta\le\sum\limits_{i=1}^3 \alpha_i -1\ri\prod\limits_{i=1}^3\alpha_i^{\nu_i-1}d\alpha_i}
{(\alpha_1\alpha_2p_3^2 
+ \alpha_2\alpha_3 p_1^2 + \alpha_1\alpha_3 p_2^2 )^{d/2}} \\
= \frac{\G(d/2)}{\prod\limits_i \G(\nu_i) } \int\limits_0^1~d\alpha_1\int\limits_0^{1-\alpha_1}d\alpha_2~\frac{\alpha_1^{\nu_1-1}\alpha_2^{\nu_2-1}(1-\alpha_1-\alpha_2)^{\nu_3-1}}{(\alpha_1\alpha_2p_3^2
+ \alpha_2(1-\alpha_1-\alpha_2) p_1^2 + \alpha_1(1-\alpha_1-\alpha_2) p_2^2 )^{d/2}} \\
= \frac{\G(d/2)}{\prod\limits_i \G(\nu_i) } \int\limits\limits_0^1~d\alpha_1\int\limits_0^{1}d\alpha_2~\frac{(1-\alpha_1)\alpha_1^{\nu_1-1}\alpha_2^{\nu_2-1}(1-\alpha_1)^{\nu_2-1} (1-\alpha_1)^{\nu_3-1} (1-\alpha_2)^{\nu_3-1}}{(\alpha_1\alpha_2 (1-\alpha_1)p_3^2
+ \alpha_2(1-\alpha_1)^2(1-\alpha_2) p_1^2 + \alpha_1(1-\alpha_1)(1-\alpha_2) p_2^2 )^{d/2}} \\
= \frac{\G(d/2)}{\prod\limits_i \G(\nu_i) } \int\limits_0^1~d\alpha_1\int\limits_0^{1}d\alpha_2~\frac{(1-\alpha_1)\alpha_1^{\nu_1-1}\alpha_2^{\nu_2-1}(1-\alpha_1)^{\nu_2-1} (1-\alpha_1)^{\nu_3-1} (1-\alpha_2)^{\nu_3-1}}
{(1-\alpha_1)^{d/2}(\alpha_1\alpha_2 p_3^2 + \alpha_2(1-\alpha_1)(1-\alpha_2) p_1^2 + \alpha_1(1-\alpha_2) p_2^2 )^{d/2}} \\
= \frac{\G(d/2)}{\prod\limits_i \G(\nu_i) } \int\limits_0^1~d\alpha_1\int\limits_0^{1}d\alpha_2~\frac{\alpha_1^{\nu_1-1}\alpha_2^{\nu_2-1}(1-\alpha_1)^{\nu_2+\nu_3-1-d/2} (1-\alpha_2)^{\nu_3-1}}
{(\alpha_1\alpha_2 p_3^2 + \alpha_2(1-\alpha_1)(1-\alpha_2) p_1^2 + \alpha_1(1-\alpha_2) p_2^2 )^{d/2}} \\
= \frac{\G(d/2)}{\prod\limits_i \G(\nu_i) } \int\limits_0^1~d\alpha_1\int\limits_0^{1}d\alpha_2~\frac{\alpha_1^{\nu_1-1}\alpha_2^{\nu_2-1}(1-\alpha_1)^{d/2-\nu_1-1} (1-\alpha_2)^{\nu_3-1}}
{(\alpha_1 (\alpha_2 p_3^2 - \alpha_2(1-\alpha_2) p_1^2 + (1-\alpha_2) p_2^2)  + \alpha_2(1-\alpha_2) p_1^2  )^{d/2}} \\
= \frac{\G(d/2)}{\prod\limits_i \G(\nu_i) } \int\limits_0^{1}d\alpha_2~\frac{\alpha_2^{\nu_2-1} (1-\alpha_2)^{\nu_3-1}{\rm B}(\nu_1,d/2-\nu_1)}
{(\alpha_2 p_3^2 + (1-\alpha_2) p_2^2))^{\nu_1}(\alpha_2(1-\alpha_2) p_1^2  )^{d/2-\nu_1}}  \\
= \frac{\G(d/2){\rm B}(\nu_1,d/2-\nu_1)}{\prod\limits_i \G(\nu_i) (p_1^2 )^{d/2-\nu_1}} \int\limits_0^{1}d\alpha_2~\frac{\alpha_2^{\nu_2-1} (1-\alpha_2)^{\nu_3-1}}
{(\alpha_2 p_3^2 + (1-\alpha_2) p_2^2))^{\nu_1}(\alpha_2(1-\alpha_2) )^{d/2-\nu_1}} \\
= \frac{\G(d/2){\rm B}(\nu_1,d/2-\nu_1)}{\prod\limits_i \G(\nu_i) (p_1^2 )^{d/2-\nu_1}} \int\limits_0^{1}d\alpha_2~\frac{\alpha_2^{\nu_2+\nu_1-1-d/2} (1-\alpha_2)^{\nu_3+\nu_1-1-d/2}}
{(\alpha_2 p_3^2 + (1-\alpha_2) p_2^2))^{\nu_1}}  \\ 
= \frac{\G(d/2){\rm B}(\nu_1,d/2-\nu_1)}{\prod\limits_i \G(\nu_i) (p_1^2 )^{d/2-\nu_1}} \int\limits_0^{1}d\alpha_2~\frac{\alpha_2^{d/2-\nu_3 -1} (1-\alpha_2)^{d/2-\nu_2-1}}
{(\alpha_2 p_3^2 + (1-\alpha_2) p_2^2))^{\nu_1}} \\
= \frac{\G(d/2){\rm B}(\nu_1,d/2-\nu_1) {\rm B}(d/2-\nu_2,d/2-\nu_3)}{\prod\limits_{i=1}^3 \G(\nu_i) (p_1^2 )^{d/2-\nu_1}(p_2^2)^{d/2-\nu_2}(p_3^2)^{d/2-\nu_3}}
= \frac{\prod\limits_{i=1}^3 \G(d/2-\nu_i)}{\prod\limits_{i=1}^3 \G(\nu_i) (p_i^2 )^{d/2-\nu_i}}.
\end{eqnarray*}
\end{adjustwidth}

We observe that $\nu_i \in ]0,d/2[,$  $i \in \{1,2,3\}$  is necessary for the convergence of the integrals that appear in this chain of transformations.  The Feynman formula should be continued analytically  for arbitrary values of $\nu_i$ but we do not need this for the moment. For us it is important that the traditional representation of the Feynman formula  (\ref{feynman-2}) in terms of the integral over simplex in the intervals $[0,1]$ is valid, that is, we are in the convergence domain of the indices $\nu_i$ in order to have this representation (\ref{triangle-FP}) valid.

Another important point is that we managed to rewrite by rescaling the integration over simplex to the integration over the unit square. The integration domain can always be transformed  to the unit square  $[0,1] \times [0,1]$  by the rescaling that we used in this example but in general it does not result in the Euler beta function. Only for this special case when the sum of the indices is $d$ this is reduced to the Euler beta functions according to the set of integrals developed in Subsection \ref{subsec-Set}. But for arbitrary indices which are not related by the uniqueness condition we may integrate the Feynman variables $\alpha_i$ out   to the Euler beta functions too, but when these integrals are part of  integrands of the Mellin-Barnes contour integrals. This approach is considered in the next section  \ref{sec-out-Mellin-Barnes}. The main idea of the method remains the same as we  used for the uniqueness case studied in this Subsection \ref{subsec-uniq}. We apply rescaling and then transform the integrals over simplex into the integrals over the unit cube.  Such integrals can be represented in terms of the Euler beta function after some auxiliary regularization which becomes necessary by the reasons we explain in the next section \ref{sec-out-Mellin-Barnes}. The uniqueness formula sometimes is called star-triangle relation. The first results on the uniqueness method  can be found   in  \cite{Unique,Vasiliev:1981dg,Vasil, Usyukina:1983gj,Kazakov:1984bw}.

\section{Integrating Feynman parameters by Mellin-Barnes transformations} \label{sec-out-Mellin-Barnes}

In this Section  \ref{sec-out-Mellin-Barnes}   we consider the integrals over Feynman parameters by applying the Mellin-Barnes transformation to the denominator of the integral representation (\ref{triangle-FP}) for the indices $\nu_1,\nu_2,\nu_3$
in the domain admitted by convergence  of the integral (\ref{triangle-FP}).  We integrate the Feynman variables $\alpha_i$ out first for the triangle integral (\ref{triangle-FP}) for the case of arbitrary indices $\nu_i$ such that Feynman formula (\ref{triangle-FP}) for them is convergent and then consider the box integral for the simple case of all the indices equal to 1 and $d=4$.

\subsection{Two-fold Mellin-Barnes integral from the triangle integral} \label{subsec-two-fold-triangle}


In this Section    we re-write formula (\ref{triangle-FP}) in terms of Barnes integral by using the
formulas of Section \ref{sec-MMellin-BarnesT} for the denominator of the formula (\ref{triangle-FP}). We obtain
\begin{eqnarray} \label{known}
J(\nu_1,\nu_2,\nu_3) = \frac{\G\le\sum\limits_{i=1}^3 \nu_i - \frac{d}{2}\ri}{\prod\limits_i \G(\nu_i) }\int\limits_0^1\int\limits_0^1\int\limits_0^1\frac{\delta\le\sum\limits_{i=1}^3 \alpha_i -1\ri\prod\limits_{i=1}^3\alpha_i^{\nu_i-1}d\alpha_i}{(\alpha_1\alpha_2p_3^2
+ \alpha_2\alpha_3 p_1^2 + \alpha_1\alpha_3 p_2^2 )^{ \sum\limits_i \nu_i - d/2}} \\
= \frac{1}{\prod\limits_i \G(\nu_i) }\int\limits_0^1\int\limits_0^1\int\limits_0^1
\frac{\delta\le\sum\limits_{i=1}^3 \alpha_i -1\ri\prod\limits_{i=1}^3\alpha_i^{\nu_i-1}d\alpha_i}{(\alpha_1\alpha_2p_3^2)^{ \sum\limits_i \nu_i - d/2}}
\int\limits_{-\delta_1-i\infty}^{-\delta_1 + i\infty} \G(-z_1)~dz_1 \int\limits_{-\delta_2-i\infty}^{-\delta_2 + i\infty}\G(-z_2) \label{Mellin-BarnesIO} \\
\G \le \sum\limits_{i= 1}^3 \nu_i - \frac{d}{2} + z_1 +z_2 \ri\le\frac{\alpha_3p_1^2}{\alpha_1p_3^2}\ri^{z_1}\le\frac{\alpha_3p_2^2}{\alpha_2p_3^2}\ri^{z_2} ~dz_2\no
\end{eqnarray}

\begin{eqnarray}
= \frac{1}{\prod\limits_i\G(\nu_i)} \frac{1}{(p_3^2)^{ \sum\limits_i \nu_i - d/2}}\int\limits_{-\delta_1-i\infty}^{-\delta_1 + i\infty} \G(-z_1)~dz_1 \int\limits_{-\delta_2-i\infty}^{-\delta_2 + i\infty}\G(-z_2)\G \le \sum\limits_{i=1}^3 \nu_i - \frac{d}{2} + z_1 +z_2 \ri \no\\
\le\frac{p_1^2}{p_3^2}\ri^{z_1}\le\frac{p_2^2}{p_3^2}\ri^{z_2} ~dz_2
\int\limits_0^1\int\limits_0^{1-\alpha_1}~d\alpha_2 d\alpha_1
\frac{\alpha_1^{\nu_1-1} \alpha_2^{\nu_2-1}(1-\alpha_1-\alpha_2)^{\nu_3-1}}{(\alpha_1\alpha_2)^{ \sum\limits_i \nu_i - d/2}} \no\\
\times \le\frac{1-\alpha_1-\alpha_2}{\alpha_1}\ri^{z_1} \le\frac{1-\alpha_1-\alpha_2}{\alpha_2}\ri^{z_2} \no \\
= \frac{1}{\prod\limits_i\G(\nu_i)} \frac{1}{(p_3^2)^{ \sum\limits_i \nu_i - d/2}}\int\limits_{-\delta_1-i\infty}^{-\delta_1 + i\infty} \G(-z_1)~dz_1 \int\limits_{-\delta_2-i\infty}^{-\delta_2 + i\infty}\G(-z_2)\G \le \sum\limits_{i=1}^3 \nu_i - \frac{d}{2} + z_1 +z_2 \ri \no\\
\le\frac{p_1^2}{p_3^2}\ri^{z_1}\le\frac{p_2^2}{p_3^2}\ri^{z_2} ~dz_2
\int\limits_0^1\int\limits_0^{1}~d\alpha_2 d\alpha_1
\frac{\alpha_1^{\nu_1-1} \alpha_2^{\nu_2-1}(1-\alpha_1)^{\nu_2+\nu_3-1}  (1-\alpha_2)^{\nu_3-1}  }{(\alpha_1\alpha_2)^{ \sum\limits_i \nu_i - d/2}(1-\alpha_1)^{ \sum\limits_i \nu_i - d/2}} \no\\
\times \le\frac{(1-\alpha_1)(1-\alpha_2)}{\alpha_1}\ri^{z_1} \le\frac{1-\alpha_2}{\alpha_2}\ri^{z_2} \no \\
= \frac{1}{\prod\limits_i\G(\nu_i)} \frac{1}{(p_3^2)^{ \sum\limits_i \nu_i - d/2}}\int\limits_{-\delta_1-i\infty}^{-\delta_1 + i\infty} \G(-z_1)~dz_1 \int\limits_{-\delta_2-i\infty}^{-\delta_2 + i\infty}\G(-z_2)\G \le \sum\limits_{i=1}^3 \nu_i - \frac{ d}{2} + z_1 +z_2 \ri \no\\
\le\frac{p_1^2}{p_3^2}\ri^{z_1}\le\frac{p_2^2}{p_3^2}\ri^{z_2} ~dz_2
\int\limits_0^1\int\limits_0^{1}~d\alpha_2 d\alpha_1
\alpha_1^{d/2 - \nu_2 -\nu_3 -z_1-1} \alpha_2^{d/2- \nu_1-\nu_3-z_2-1} \no\\
\times (1-\alpha_1)^{d/2-\nu_1 + z_1-1}  (1-\alpha_2)^{z_1+z_2+ \nu_3-1} \label{aqui}  \\
= \frac{1}{\prod\limits_i\G(\nu_i)} \frac{1}{(p_3^2)^{ \sum\limits_i \nu_i - d/2}}\int\limits_{-\delta_1-i\infty}^{-\delta_1 + i\infty} \G(-z_1)~dz_1 \int\limits_{-\delta_2-i\infty}^{-\delta_2 + i\infty}\G(-z_2)\G \le \sum\limits_{i=1}^3 \nu_i - \frac{d}{2} + z_1 +z_2 \ri  \no\\
\le\frac{p_1^2}{p_3^2}\ri^{z_1}\le\frac{p_2^2}{p_3^2}\ri^{z_2} ~dz_2
\frac{\G\le \frac{d}{2} - \nu_2 -\nu_3 -z_1\ri\G\le  \frac{d}{2}-\nu_1+z_1\ri}{\G\le d -   \sum\limits_{i=1}^3 \nu_i \ri} \no\\
\times\frac{\G\le \frac{d}{2}- \nu_1-\nu_3-z_2\ri \G\le z_1+z_2+ \nu_3\ri}
{\G\le \frac{d}{2} - \nu_1 +z_1  \ri} \label{known-final}
\end{eqnarray}
The same result was used in Ref.\cite{Allendes:2012mr,Gonzalez:2016pgx} from which it had been proved in Ref.\cite{Gonzalez:2016pgx} that the reduction of the loop number is equivalent to the first and second Barnes lemmas.

Looking at the first line of this expression (\ref{known})  we can observe that the positivity of the real parts of the indices $\nu_i$ is necessary for convergence of  (\ref{known}).   Looking at the line (\ref{aqui}) we can conclude that for convergence it is also necessary that the sum of any of two indices from the set $\nu_1,\nu_2,\nu_3$ is less than $d/2.$ The integral (\ref{known}) cannot be analytically continued to arbitrary $\nu_i$ outside of the convergence domain without a necessary modification of the Feynman formula. Taking into account the values of the ${\rm Re}~z_1$ and   ${\rm Re}~z_2$ in the Mellin-Barnes integrals (\ref{two-fold-1}, \ref{Mellin-BarnesIO})
over the straight lines we can conclude that these real parts of the Mellin variables do not spoil
the convergence of the integrals over the Feynman parameters in the line (\ref{aqui}).

In Subsection \ref{subsec-five} when we analyze the box diagram the analytic regularization of the integrals over Feynman variables in the integrand of the Mellin-Barnes integrals is possible because the values of the indices in the case of simple box   belong to the domain of convergence and we know that the overall simple box integral is finite.   It means that a regularization for the Mellin-Barnes integrands is admitted.  We can shift the powers of the Feynman parameters in the Mellin-Barnes integrands by sufficiently big numbers to guarantee the convergence of the integrals over the Feynman parameters, take the integral over Feynman parameters off, and then tend these values of the shift to zero.  A very similar trick has been used in Refs. \cite{Faouzi,Gonzalez:2023jig} for other types of integrals.  As we will see in Section \ref{subsec-five} the divergence of the integrals over Feynman variables  makes situation even simpler than in the regular case when these integrals over the Feynman variables are well-defined.

\subsection{Two-fold Mellin-Barnes  integral for $d=4$ and $\nu_1=\nu_2=\nu_3=1$}

In this case the two-fold Mellin-Barnes representation is simple. It follows from Eq. (\ref{known-final})
\begin{eqnarray*}
J(1,1,1) = \frac{1}{p^2_3}\int\limits_{-\delta_1-i\infty}^{-\delta_1+i\infty}    
dz_1\int\limits_{-\delta_2-i\infty}^{-\delta_2+i\infty}dz_2 \le\frac{p^2_1}{p^2_3}\ri^{z_1}  \le\frac{p^2_2}{p^2_3}\ri^{z_2}
\G^2 \le -z_1 \ri \G^2 \le -z_2 \ri \G^2 \le 1 + z_1 + z_2 \ri
\end{eqnarray*}

\subsection{Integrating three and four Feynman parameters out} \label{subsec-3-4}


Let us suppose for this Subsection (\ref{subsec-3-4})  that an integral over the Feynman variables is convergent at the upper and lower limits. We suppose that such an integral is a part of the integrand in the Mellin-Barnes integral, for a example as it was in Subsection  \ref{subsec-two-fold-triangle}. We consider first
the integration over a simplex for the case of three Feynman variables
\begin{eqnarray} \label{block-3}
\int\limits_0^1\int\limits_0^1\int\limits_0^1\delta\le\sum_{i=1}^{3} \alpha_i -1\ri~
\alpha_1^{z}\alpha_2^{u}\alpha_3^{v}\prod_{i=1}^3 d\alpha_i  = \int\limits_0^1 d\alpha_1 \int\limits_0^{1-\alpha_1} d\alpha_2~\alpha_1^{z}\alpha_2^{u}(1-\alpha_1 - \alpha_2)^{v} \no\\
 = \int\limits_0^1 d\alpha_1 \int\limits_0^1 d\alpha_2  ~\alpha_1^{z}\alpha_2^{u} (1-\alpha_1)^{u+v+1}(1-\alpha_2)^v
 \no\\
 = \int\limits_0^1 d\alpha_1   ~\alpha_1^{z}(1-\alpha_1)^{u+v+1}
 \int\limits_0^1 d\alpha_2  ~\alpha_2^{u} (1-\alpha_2)^v  = {\rm B}(z+1,u+v+2) ~{\rm B}(u+1,v+1)
 \end{eqnarray}
This simple chain of transformations has been used in Subsection \ref{subsec-two-fold-triangle}. The most important point here is that we can transform the integral over simplex into the integrals over the unit square  $[0,1]\times[0,1]$  as in (\ref{block-3})
or over  the  unit cube  $[0,1]\times[0,1]\times[0,1] $  as in (\ref{block-4}),
 which can be easily transformed to the Euler beta functions if these integrals
converge. We suppose that the set of the complex powers $z,u,v$ which parametrize this integral  (\ref{block-3})
   belongs to the convergence domain of the indices in the Feynman formula.

Now we repeat the same chain of the transformation for the integration over simplex with four Feynman
variables. Again, we suppose that all the complex indices of the set $z,u,v,w$ which appear in the integral over simplex with four Feynman variables (\ref{block-4})  belong to the domain of convergence of this integral   (\ref{block-4}), that is the integral (\ref{block-4}) is convergent at the upper and lower limits. We obtain
\begin{eqnarray} \label{block-4}
\int\limits_0^1\int\limits_0^1\int\limits_0^1\int\limits_0^1\delta\le\sum_{i=1}^{4} \alpha_i -1\ri~
\alpha_1^{z}\alpha_2^{u}\alpha_3^{v}\alpha_4^{w}\prod_{i=1}^4 d\alpha_i \\
= \int\limits_0^1 d\alpha_4 \int\limits_0^{1-\alpha_4} d\alpha_3 \int\limits_0^{1- \alpha_4 - \alpha_3}d\alpha_2
 ~(1-\alpha_2 - \alpha_3 - \alpha_4)^{z}\alpha_2^{u}\alpha_3^{v}\alpha_4^{w} \no\\
 = \int\limits_0^1 d\alpha_4 \int\limits_0^1 d\beta_3 \int\limits_0^{1- \beta_3}d\beta_2 (1-\alpha_4)^2
 ~(1-\beta_2 - \beta_3)^{z}(1-\alpha_4)^{z+u+v}  \beta_2^{u}\beta_3^{v}\alpha_4^{w} \no\\
 = \int\limits_0^1 d\alpha_4  \alpha_4^{w}(1-\alpha_4)^{2+z+u+v}
   \int\limits_0^1 d\beta_3 \int\limits_0^{1- \beta_3}d\beta_2
   ~(1-\beta_2 - \beta_3)^{z}  \beta_2^{u}\beta_3^{v} \no\\
= \int\limits_0^1 d\alpha_4  \alpha_4^{w}(1-\alpha_4)^{2+z+u+v}
   \int\limits_0^1 d\beta_3 \int\limits_0^{1}d\gamma_2 (1- \beta_3)
   ~(1-\gamma_2)^{z} (1- \beta_3)^{z+u} \gamma_2^{u}\beta_3^{v} \no\\
= \int\limits_0^1 d\alpha_4  \alpha_4^{w}(1-\alpha_4)^{2+z+u+v}
   \int\limits_0^1 d\beta_3 (1- \beta_3)^{1+z+u} \beta_3^{v}  \int\limits_0^{1}d\gamma_2
   ~(1-\gamma_2)^{z}  \gamma_2^{u} \no
\end{eqnarray}
In Subsection \ref{subsec-five} we have to consider the case when the set of the indices  $z,u,v,w$ in the integral (\ref{block-4}) does not belong to the convergence domain. This happens when such integrals over the Feynman variables are parts of integrands of Mellin-Barnes contours integrals. In such a case an auxiliary regularization is necessary to make this integrals convergent. This regularization will be removed at the end. Examples of application of such a regularization in mathematical statistics and statistical mechanics may be found in \cite{Gonzalez:2023jig,Faouzi}. With such a regularization the divergent integrals of this type (\ref{block-4}) can be represented in terms of the Euler beta functions.

\subsection{Box integral over Feynman parameters}

\begin{figure}[H] 
\centering\includegraphics[scale=0.3]{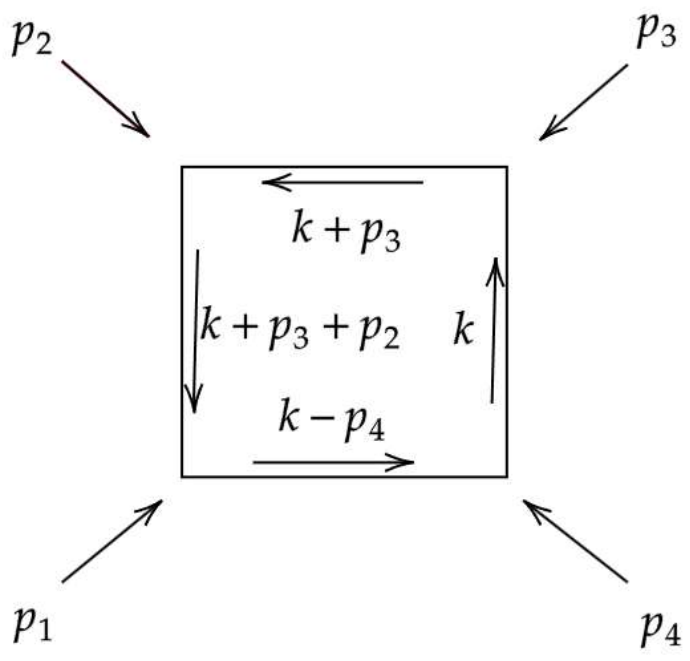}
\caption{\footnotesize  One-loop massless scalar triangle in momentum space}
\label{figure-2}
\end{figure}

The massless box diagram is depicted in this Fig. \ref{figure-2}

This box integral over Feynman parameters
\begin{eqnarray}  \label{Box-4D-alfa}
\int\limits_{0}^{1}\int\limits_{0}^{1}\int\limits_{0}^{1}\int\limits_{0}^{1}    \frac{\delta\le\sum\limits_{i=1}^{4} \alpha_i -1\ri\prod\limits_{i=1}^4 d\alpha_i}
{[\alpha_1\alpha_2p_1^2 + \alpha_2\alpha_3 p_2^2 + \alpha_3\alpha_4 p_3^2 + \alpha_1\alpha_4p_4^2 + \alpha_1\alpha_3 s + \alpha_2\alpha_4 t ]^2}
\end{eqnarray}
corresponds to all the indices 1 and $d=4.$

\subsection{From the box to the triangle via Feynman parameters}

In the paper \cite{Usyukina:1992jd} it has been shown that the box diagram and the triangle diagram are equivalent
under the condition that all the indices of the propagators  are 1 and the dimension of the spacetime is 4 exactly.  In Appendix
\ref{App} we give a brief review of the trick  used in  \cite{Usyukina:1992jd}  which suggests a combination of the Feynman formula  and
Mellin-Barnes integral representation. We reproduce it  just to show that we may avoid  extensive mixture of Feynman formula and Mellin-Barnes
transformation and may work only in terms of Mellin-Barnes transformation from the very beginning. In essence, we compare pure Mellin-Barnes approach and the approach of Usyukina and Davydychev.  We just would like to notice that the pure Mellin-Barnes approach is necessary  to elaborate the strategy in order to reduce the number of contour
integrals what basically is equivalent to reduction of the rank of the corresponding series resulting from evaluation of the residues by Cauchy integral formula.  Reduction of the rank of the resulting special functions means that some of the contours contribute with a finite number of residues
that is highly unusual for the Barnes integrals because the Euler gamma functions  produce infinite number of residues. 
Any Barnes integral can be represented in terms of special functions,  no matter if the integrals over Feynman parameters contribute with a finite number of residues or with an infinite number of residues in the complex planes of the corresponding complex integrals.

In the next Subsection \ref{subsec-five}
of the present article we show that the Mellin-Barnes representation of (\ref{Box-4D-alfa})  is enough to reduce the
number of independent arguments in the resulting hypergeometry, that is, to reduce the rank of the resulting hypergeometry and it is not necessary to use Feynman formula at an intermediate step. We map the Feynman formula (\ref{Box-4D-alfa}) to the Mellin-Barnes representation at the initial step, and then never go back, all the work is done by the analytical regularization of the divergent integrals over Feynman parameters
in the integrand of the contour Mellin-Barnes integrals.

Also, we need to comment that the trick of \cite{Usyukina:1992jd} is not the unique way to transform the triangle ladder diagram to the box ladder diagram. The simplest way is to apply conformal transformation in an auxiliary
position space which is dual to the momentum space \cite{Kondrashuk:2009us,Broadhurst:1993ib}

\subsection{The five-fold Mellin-Barnes box is the two-fold Mellin-Barnes triangle} \label{subsec-five}

Our task is to reproduce the same result (\ref{UD1})  directly from the integral (\ref{Box-4D-alfa}) which is an integral over Feynman parameters.
This will help us to elaborate a strategy how to reduce the number of contour integration in the Mellin-Barnes representations  of the integrals over Feynman  parameters in a more general case  of arbitrary spacetime dimension and arbitrary indices of the propagators.  The strategy that is formulated in the next Section  \ref{sec-Discuss} is based on the experience we gain in this Section \ref{subsec-five}. Advantage of such an approach is that the calculation via Mellin-Barnes works everywhere while  the trick of Usyukina and Davydychev described in \ref{App}  works only for  very special cases when the indices of the propagators and the dimension of spacetime are related.  The integrals over Feynman  parameters result in product of the Euler beta functions via the algorithm we have outlined in Subsection \ref{subsec-3-4}, however it may happen that an additional regularization of singularities  is required in order to represent such integrals in terms of the Euler beta functions. Remarkably, when
such singularities appear in the integration over the straight lines in the Barnes integrals, the
number of the integration contours in the corresponding complex domains may be  reduced.

We show in this section  \ref{subsec-five} how the multi-fold Mellin-Barnes integrals  can be reduced to the  two-fold Mellin-Barnes integrals. The multi-fold Mellin-Barnes integrals we consider here are produced from the integrals over Feynman parameters like for example (\ref{Box-4D-alfa}).  The reason for such a reduction  is in an efficient regularization of the divergent integrals over Feynman parameters.   We observe that each integral over Feynman parameters regularized analytically  produces infinite number of residues (as usual for the Euler beta function)   but only one of them contributes in the limit of removing the analytical regularization.   These residues  are situated in certain complex planes
which correspond to certain linear combinations of the Mellin-Barnes integration variables.    When the sum of the powers of the Feynman parameters after the Mellin-Barnes transformation of  denominators is a non-positive integer, a rearrangement of the Mellin-Barnes integration variables
in their complex planes  is necessary.  Below we explain how to establish which combination of Mellin-Barnes variables contributes with a finite number of residues.  The example considered in this Subsection  \ref{subsec-five} explains how to rearrange the variables in order to reduce maximally the number  of the integration contours.

First of all, this is the five-fold Mellin-Barnes transformation because of the six terms in the denominator
\begin{eqnarray*} 
\int\limits_0^1\int\limits_0^1\int\limits_0^1\int\limits_0^1 \frac{\delta\le\sum\limits_{i=1}^{4} \alpha_i -1\ri\prod\limits_{i=1}^4 d\alpha_i}
{(\alpha_1\alpha_2 p_1^2 + \alpha_2\alpha_3 p_2^2 + \alpha_3\alpha_4 p_3^2 + \alpha_1\alpha_4 p_4^2 + \alpha_1\alpha_3 s + \alpha_2\alpha_4 t )^2} \no\\
= 
\int\limits_{-\delta_1 -i\infty}^{-\delta_1 + i\infty} ~dz_1\int\limits_{-\delta_2 -i\infty}^{-\delta_2  + i\infty} ~dz_2\int\limits_{-\delta_3 -i\infty}^{-\delta_3  + i\infty} ~dz_3\int\limits_{-\delta_4 -i\infty}^{-\delta_4 + i\infty} ~dz_4\int\limits_{-\delta_5-i\infty}^{-\delta_5 + i\infty} ~dz_5\G(-z_1)\G(-z_2) \no\G(-z_3)\G(-z_4)\\
 \times 
\G(-z_5)\G(2+z_1+z_2+z_3+z_4+z_5)\int\limits_0^1\int\limits_0^1\int\limits_0^1\int\limits_0^1
\frac{ \delta\le\sum\limits_{i=1}^{4} \alpha_i -1\ri\prod\limits_{i=1}^4 d\alpha_i  }{(\alpha_1\alpha_2p_1^2)^2} \no\\
\times\le\frac{\alpha_2\alpha_3p_2^2}{\alpha_1\alpha_2p_1^2}\ri^{z_1}\le\frac{\alpha_3\alpha_4p_3^2}{\alpha_1\alpha_2p_1^2}\ri^{z_2}
\le\frac{\alpha_1\alpha_4p_4^2}{\alpha_1\alpha_2p_1^2}\ri^{z_3}
\le\frac{\alpha_1\alpha_3s}{\alpha_1\alpha_2p_1^2}\ri^{z_4}
\le\frac{\alpha_2\alpha_4t}{\alpha_1\alpha_2p_1^2}\ri^{z_5} \no\\
\no\\
= \frac{1}{(p_1^2)^2}\int\limits_{-\delta_1 -i\infty}^{-\delta_1 + i\infty} ~dz_1\int\limits_{-\delta_2 -i\infty}^{-\delta_2  + i\infty} ~dz_2\int\limits_{-\delta_3 -i\infty}^{-\delta_3 + i\infty} ~dz_3\int\limits_{-\delta_4 -i\infty}^{-\delta_4  + i\infty} ~dz_4\int\limits_{-\delta_5-i\infty}^{-\delta_5 + i\infty} ~dz_5\G(-z_1)\G(-z_2)
\end{eqnarray*}
\begin{eqnarray} \label{five}
\times\G(-z_3)\G(-z_4)\G(-z_5)\G(2+z_1+z_2+z_3+z_4+z_5)
\le\frac{p_2^2}{p_1^2}\ri^{z_1}
\le\frac{p_3^2}{p_1^2}\ri^{z_2}\le\frac{p_4^2}{p_1^2}\ri^{z_3}
\le\frac{s}{p_1^2}\ri^{z_4}\no\\
\le\frac{t}{p_1^2}\ri^{z_5}
\int\limits_0^1\int\limits_0^1\int\limits_0^1\int\limits_0^1
 \delta\le\sum_{i=1}^{4} \alpha_i -1\ri~
\alpha_1^{-2-z_1-z_2-z_5}
\alpha_2^{-2-z_2-z_3-z_4}\alpha_3^{z_1+z_2+z_4}\alpha_4^{z_2+z_3+z_5}\prod_{i=1}^4 d\alpha_i
\end{eqnarray}
The hierarchy of the contour positions in the multi-fold Mellin-Barnes transformation considered in Section \ref{sec-MMellin-BarnesT} suggests
\begin{eqnarray*}
{\rm Re}~\lambda  > 0  \\
{\rm Re}~\lambda > \delta_1 > 0  \\
{\rm Re}~\lambda-\delta_1 > \delta_2 > 0 \\
{\rm Re}~\lambda-\delta_1  - \delta_2 > \delta_3 > 0 \\
{\rm Re}~\lambda-\delta_1  - \delta_2 - \delta_3 >  \delta_4 > 0 \\
{\rm Re}~\lambda-\delta_1  - \delta_2 - \delta_3 -  \delta_4 >  \delta_5 >0 \\
\end{eqnarray*}

  We converted the integration over the simplex into the integration over the unit cube in the way established in Subsection \ref{subsec-3-4}. This is a universal step in order to take the integration over the Feynman parameters off.  An intermediate analytical regularization may be necessary to generate certain ratios of the Euler gamma functions. In Subsection \ref{subsec-two-fold-triangle} we have not used any analytical intermediate regularization supposing that the indices  belong to the domain of the convergence at both the limits, upper and lower. The result is a Barnes integral in the integrand of which there is  a product of the gamma functions divided by another product of the gamma functions. In case when $\G(-n+\varepsilon)$ in the denominator
is generated,  $\varepsilon$ is a complex parameter of the analytic regularization,  the integration becomes simple because only a few residues  contribute to the limit
$\varepsilon \rightarrow 0$ in which we remove the intermediate analytical regularization.
A similar situation  can be discovered  for the uniqueness case \cite{Gonzalez:2018gch} however the $\G(0)$ has appeared in the denominator by construction due to the uniqueness condition but not due to the integration over Feynman parameters as we obtain here.  In the integral (\ref{five}) that we consider in the present
paper the structure of the Mellin-Barnes integrands repeats the structure of the integrands used to prove the orthogonality of triangles in \cite{Gonzalez:2018gch}.  This fact suggests to rearrange variables of Mellin-Barnes integration in order to make it explicit  which combinations of the initial complex variables  should be integrated out in the first few steps.

Suppose that we intent to convert the integral over the simplex in the last line of Eq. (\ref{five})
 into the integral over the unit
cube by using Eq. (\ref{block-4}) from Subsection \ref{subsec-3-4}. However, this integral over simplex is divergent for each of the  powers of the Feynman variables which are given in this last line of (\ref{five}).
The natural question may be asked. Namely,
can we make a change of variables in a  divergent integral over a simplex with these given powers of Feynman variables? The real parts of these complex powers are fixed by the positions of the straight vertical lines in the integration contours of the complex domains of the Mellin-Barnes integrals for the inverse transformation  in (\ref{five}).
The powers of the Feynman parameters in the last line of (\ref{five})  are integrated over five contours with another function of these powers
from the integrand of the contour integral and this overall integral in five complex domains is convergent.
In turn, the convergence of this overall integral over five complex contours means that the last line of
(\ref{five})  should be understood in a sense of distributions and a  change of variables for the multiple integral over three Feynman parameters can be done.   We  introduce the analytical regularization at the level of the integral  (\ref{five})  beginning  with shifts  in the  
 powers of the Feynman parameters  in the integrand over the simplex  in the last line of (\ref{five})  by specific 
 combinations of  complex parameters  $\varepsilon_1,\varepsilon_2,\varepsilon_3$   of the analytical regularization  which serve to regularize the integrations over the simplex  and  to guarantee the swapping between the Riemann integrations over Feynman parameters and the Mellin-Barnes integrations over $z_1, z_2, z_3,  z_4, z_5.$ Consequently, we introduce   a product of certain powers  of the Feynman parameters $\alpha_1,\alpha_2,\alpha_3,\alpha_4$ within the numerator of the first integral in the chain (\ref{five}) of equalities.   These powers  depend solely  on the regularization parameters    $\varepsilon_1,\varepsilon_2,\varepsilon_3.$ 
 By changing the simplex to the unit cube we obtain

\begin{eqnarray*} 
\int\limits_0^1\int\limits_0^1\int\limits_0^1\int\limits_0^1
 \delta\le\sum_{i=1}^{4} \alpha_i -1\ri~
\alpha_1^{-2-z_1-z_2-z_5}
\alpha_2^{-2-z_2-z_3-z_4}\alpha_3^{z_1+z_2+z_4}\alpha_4^{z_2+z_3+z_5}\prod_{i=1}^4 d\alpha_i \no\\
= \int\limits_0^1~ d\alpha_4 ~ \alpha_4^{z_2+z_3+z_5}(1-\alpha_4)^{-2-z_5-z_2-z_3}
   \int\limits_0^1 d\beta_3 (1- \beta_3)^{-3-z_1-2z_2-z_3-z_4-z_5} \beta_3^{z_1+z_2+z_4} \no
\end{eqnarray*}
\begin{eqnarray} \label{3-beta}
\times   \int\limits_0^{1}d\gamma_2 ~(1-\gamma_2)^{-2-z_1-z_2-z_5}  \gamma_2^{-2-z_2-z_3-z_4}.
\end{eqnarray}
This means that in the formula  (\ref{block-4})  the powers of the Feynman variables are chosen to take these values
\begin{eqnarray*}
z &=& -2-z_1-z_2-z_5 \\
u &=& -2-z_2-z_3-z_4 \\
v &=& z_1+z_2+z_4 \\
w &=& z_2+z_3+z_5,
\end{eqnarray*}

The powers of Feynman parameters  were supposed to belong to the convergent domain in formula (\ref{block-4}). However, in Eq. (\ref{3-beta}) we cannot suppose the same  because the powers are not in the domain  of convergence. As we have explained in the previous paragraph,  we still may change the integration domain for the triple integral over Feynman parameters from the simplex to the unit cube because of the overall convergence of this
five-fold Mellin-Barnes integral (\ref{five}).  As the result,  all these three integrals on the rhs of of (\ref{3-beta})
are divergent for the powers situated at  the positions of the straight vertical lines in the expression  (\ref{five}).
Because the initial integral over Feynman parameters (\ref{Box-4D-alfa}) is convergent, the integral over a simplex can be separated in a product of three integrals  even in this case when they are not convergent for such positions of the contours. 
The convergence of the overall integral suggests that we can shift the powers in these
three integrals in such a way that each of them becomes convergent and equal to the Euler beta function.  
Indeed, as it is stated in the following paragraphs of  Subsection \ref{subsec-five},  to regularize the integrals over the unit cube on the rhs  of   (\ref{3-beta}) it is sufficient to shift the power of  $ (1-\alpha_4)$  by  $\varepsilon_1 $ ( such that  ${\rm Re~~}\varepsilon_1 > 2$),   the power of  $\beta_3$ by $\varepsilon_2$  (such that  ${\rm Re~~}\varepsilon_2 > 2$)    and the power of $ (1-\gamma_2)$ by  $\varepsilon_3$ 
(such that  ${\rm Re~~}\varepsilon_3 > 1$).  This modification of the exponents  in the Riemann integrations over the unit cube
is necessary to ensure the rhs of   (\ref{3-beta})  is convergent.  Consequently,  the powers in the integrand over the simplex on the lhs of    (\ref{3-beta})  must be changed  in the corresponding manner
to reflect this modification of the powers on the rhs of (\ref{3-beta}).  Such an analytical regularization by shifting the powers in the integrand allows to swap the Riemann  integral over the simplex with the Mellin-Barnes  integrals. Furthermore,  it enables a transformation  of the integration domain in (\ref{3-beta})   from the simplex to the unit cube in a controlled  manner  ensuring  the convergence of all  integrals.
A similar idea has been used in \cite{Gonzalez:2023jig,Faouzi}.
Then, under this auxiliary analytical regularization the three resulting Euler beta functions
on the r.h.s. of (\ref{3-beta}) should be understood in the limit of the removing this auxiliary regularization
 \cite{Gonzalez:2023jig,Faouzi} as distributions which are integrated with the rest of the complex integrand over all the five contours. Then, the residues  can be calculated according to the integral Cauchy formula. We show that 
only a finite number of residues survives in this limit of the removing regularization. This happens due to the singular behavior of these three unregularized integrals on the rhs of (\ref{3-beta}).  It is singular because  the powers of the Feynman variables belong to the straight vertical lines of the Mellin-Barnes integration  contours.

Now we recombine the complex integration variables, taking into account of course the simple
structure (\ref{UD1}) we ``search''   in order to prove it 
\begin{eqnarray*} 
\frac{1}{(p_1^2)^2}\int\limits_{-\delta_1 -i\infty}^{-\delta_1  + i\infty} ~dz_1\int\limits_{-\delta_2 -i\infty}^{-\delta_2  + i\infty} ~dz_2\int\limits_{-\delta_3 -i\infty}^{-\delta_3  + i\infty} ~dz_3\int\limits_{-\delta_4 -i\infty}^{-\delta_4  + i\infty} ~dz_4\int\limits_{-\delta_5-i\infty}^{-\delta_5 + i\infty} ~dz_5\G(-z_1)\G(-z_2) \G(-z_3)\no\\
\times\G(-z_4)\G(-z_5)\G(2+z_1+z_2+z_3+z_4+z_5)
\le\frac{p_2^2}{p_1^2}\ri^{z_1}
\le\frac{p_3^2}{p_1^2}\ri^{z_2}\le\frac{p_4^2}{p_1^2}\ri^{z_3}
\le\frac{s}{p_1^2}\ri^{z_4}\no\\
\le\frac{t}{p_1^2}\ri^{z_5}
\int\limits_0^1\int\limits_0^1\int\limits_0^1\int\limits_0^1
 \delta\le\sum_{i=1}^{4} \alpha_i -1\ri~
\alpha_1^{-2-z_1-z_2-z_5}
\alpha_2^{-2-z_2-z_3-z_4}\alpha_3^{z_1+z_2+z_4}\alpha_4^{z_2+z_3+z_5}\prod_{i=1}^4 d\alpha_i \\
= \frac{1}{(p_1^2)^2}\int\limits_{-\delta_1 -i\infty}^{-\delta_1  + i\infty} ~dz_1\int\limits_{-\delta_2 -i\infty}^{-\delta_2  + i\infty} ~dz_2\int\limits_{-\delta_3 -i\infty}^{-\delta_3  + i\infty} ~dz_3\int\limits_{-\delta_4 -i\infty}^{-\delta_4  + i\infty} ~dz_4\int\limits_{-\delta_5-i\infty}^{-\delta_5 + i\infty} ~dz_5\G(-z_1)\G(-z_2)\G(-z_3) \no\\
 \no\\
\times\G(-z_4)\G(-z_5)\G(2+z_1+z_2+z_3+z_4+z_5)
\le\frac{p_2^2}{p_1^2}\ri^{z_1}
\le\frac{p_3^2}{p_1^2}\ri^{z_2}\le\frac{p_4^2}{p_1^2}\ri^{z_3}
\le\frac{s}{p_1^2}\ri^{z_4}\le\frac{t}{p_1^2}\ri^{z_5} \no\\
\int\limits_0^1~ d\alpha_4 ~ \alpha_4^{z_2+z_3+z_5}(1-\alpha_4)^{-2-z_5-z_2-z_3}
   \int\limits_0^1 d\beta_3 (1- \beta_3)^{-3-z_1-2z_2-z_3-z_4-z_5} \beta_3^{z_1+z_2+z_4} \no\\
   \int\limits_0^{1}d\gamma_2 ~(1-\gamma_2)^{-2-z_1-z_2-z_5}  \gamma_2^{-2-z_2-z_3-z_4} \no\\
 =   \frac{1}{st}\int\limits_{-\delta_1 -i\infty}^{-\delta_1  + i\infty} ~dz_1\int\limits_{-\delta_2 -i\infty}^{-\delta_2  + i\infty} ~dz_2\int\limits_{-\delta_3 -i\infty}^{-\delta_3  + i\infty} ~dz_3\int\limits_{-\delta_4 -i\infty}^{-\delta_4 + i\infty} ~dz_4\int\limits_{-\delta_5-i\infty}^{-\delta_5 + i\infty} ~dz_5\G(-z_1)\G(-z_2) \no\\
\times\G(-z_3)\G(-z_4)\G(-z_5)\G(2+z_1+z_2+z_3+z_4+z_5)
\le\frac{p_2^2 p_4^2}{st}\ri^{z_1}
\le\frac{p_3^2p_1^2}{st}\ri^{z_2}\no\\
s^{z_1+z_2+z_4+1} t^{z_1+z_2+z_5+1}
\le p_4^2\ri^{-z_1+z_3} \le p_1^2\ri^{-z_1-2z_2-z_3-z_4-z_5-2} \no \\
\int\limits_0^1~ d\alpha_4 ~ \alpha_4^{z_2+z_3+z_5}(1-\alpha_4)^{-2-z_5-z_2-z_3}
   \int\limits_0^1 d\beta_3 (1- \beta_3)^{-3-z_1-2z_2-z_3-z_4-z_5} \beta_3^{z_1+z_2+z_4} \no\\
   \int\limits_0^{1}d\gamma_2 ~(1-\gamma_2)^{-2-z_1-z_2-z_5}  \gamma_2^{-2-z_2-z_3-z_4} \no\\
=   \frac{1}{st}\int\limits_{-\delta_1 -i\infty}^{-\delta_1+ i\infty} ~dz_1\G(-z_1)\int\limits_{-\delta_2-i\infty}^{-\delta_2 + i\infty} ~dz_2 ~\G(-z_2)\int\limits_{1-\delta_2-\delta_2 -\delta_5-i\infty}^{1-\delta_1-\delta_2 -\delta_5  + i\infty} ~du~\G(1+z_1+z_2-u) \no\\
\int\limits_{\delta_1 -\delta_3 -i\infty}^{\delta_1 -\delta_3 +  i\infty} ~dv~\G(-v-z_1)\int\limits_{1-\delta_1 -\delta_2 -\delta_4 -i\infty}^{ 1-\delta_1-\delta_2 -\delta_4 + i\infty} ~dw~\G(1+z_1+z_2-w)\G(u+w+v-z_2)\no
\end{eqnarray*}

\begin{eqnarray} \label{five-2}
\le\frac{p_2^2 p_4^2}{st}\ri^{z_1}
\le\frac{p_3^2p_1^2}{st}\ri^{z_2}s^{w}t^u\le p_4^2\ri^{v} \le p_1^2\ri^{-v-w-u} \no\\
\int\limits_0^1~ d\alpha_4 ~ \alpha_4^{v+u-1}(1-\alpha_4)^{-1-u-v}
\int\limits_0^1 d\beta_3 (1- \beta_3)^{-v-w-u-1} \beta_3^{w-1} \int\limits_0^{1}d\gamma_2 ~(1-\gamma_2)^{-u-1}  \gamma_2^{-1-v-w}
\end{eqnarray}
Here we have introduced new complex variables
\begin{eqnarray} \label{position}
w &=& z_1 + z_2 + z_4 +1 \no\\
u &=& z_1+ z_2 + z_5 +1 \\
v &=& -z_1+z_3 \no
\end{eqnarray}
As we have mentioned in the previous paragraphs of this Subsection \ref{subsec-five}, exactly these combinations of the initial Mellin-Barnes integration variables can be integrated out by the Cauchy integral formula with a finite number of residues. In this particular
case each of them is integrated out with one residue only.  The rest of the infinite number of residues does snot contribute for any of these three variables $u,v,w.$   

In order to analyze convergence of the integrals over the Feynman parameters in the last line of  (\ref{five-2}) at the integration limits at the points 1 and 0, it is necessary to take into account the values of the real parts of these new variables $u,w,z$ given by the positions of the straight vertical lines of the contours of the initial Mellin-Barnes integration variables.  The integral over $\alpha_4$   in the last line of  (\ref{five-2}) is not convergent at its upper limit for the values of $u$ and $v$ on the straight vertical  lines of the  $u$ and $v$ integration contours. However, we know that the overall five-fold integral  (\ref{five-2}) (this is the same structure (\ref{five}) but in terms of new variables) is finite  because the original integral
(\ref{Box-4D-alfa}) over the Feynman parameters  is finite. This means that
\begin{eqnarray} \label{alpha}
\int\limits_0^1~ d\alpha_4 ~ \alpha_4^{v+u-1}(1-\alpha_4)^{-1-u-v}
\end{eqnarray}
should be understood in a sense of distributions over the new complex variables $u$ and $v,$  that is, the integral   (\ref{five-2})   can be concisely written as   
\begin{eqnarray}  \label{five-alpha}\frac{1}{st}
\int\limits_{1-\delta_1-\delta_2 -\delta_5-i\infty}^{1-\delta_1-\delta_2 -\delta_5  + i\infty} ~du
\int\limits_{\delta_1 -\delta_3 -i\infty}^{\delta_1 -\delta_3 +  i\infty} ~dv
~f_1(u,v)~\int\limits_0^1~ d\alpha_4 ~ \alpha_4^{v+u-1}(1-\alpha_4)^{-1-u-v},
\end{eqnarray}
$f_1(u,v)$ here stands for the rest of the integrand in (\ref{five-2}).   The  five-fold integral   (\ref{five-2}) is not singular, it takes a finite value. 
This means that  (\ref{alpha})  can be replaced in (\ref{five-alpha}) with a  limit  which regularizes  (\ref{alpha}), that is,
\begin{eqnarray*}  
 \lim_{\varepsilon_1 \rightarrow 0} \int\limits_{1-\delta_1-\delta_2 -\delta_5-i\infty}^{1-\delta_1-\delta_2 -\delta_5  + i\infty} ~du
\int\limits_{\delta_1 -\delta_3 -i\infty}^{\delta_1 -\delta_3 +  i\infty} ~dv
~f_1(u,v)~\int\limits_0^1~ d\alpha_4 ~ \alpha_4^{v+u-1}(1-\alpha_4)^{-1-u-v+\varepsilon_1} \\
 = \lim_{\varepsilon_1 \rightarrow 0} \int\limits_{1-\delta_1-\delta_2 -\delta_5-i\infty}^{1-\delta_1-\delta_2 -\delta_5  + i\infty} ~du
\int\limits_{\delta_1 -\delta_3 -i\infty}^{\delta_1 -\delta_3 +  i\infty} ~dv
~f_1(u,v)~\frac{\G(v+u)\G(-u-v+\varepsilon_1)}{\G(\varepsilon_1)} \\
= 
\int\limits_{\delta_1 -\delta_3 -i\infty}^{\delta_1 -\delta_3 +  i\infty} ~dv
~f_1(0,v),
\end{eqnarray*}
where we suppose that we have started with ${\rm Re~~}\varepsilon_1 > 2$.    This trick has been applied  in \cite{Faouzi,Gonzalez:2023jig}. Apparently, the pole $u= \varepsilon_1 -v$ in the complex $u$ plane contributes because it is on the right hand side of the straight vertical line in the complex plane of the variable $u$ when
${\rm Re~~}\varepsilon_1 > 2,$ 
\begin{eqnarray*}
\res_{u= \varepsilon_1 -v} \frac{\G(v+u)\G(-u-v+\varepsilon_1)}{\G(\varepsilon_1)} =
\res_{u= \varepsilon_1 -v} \frac{\G(v+u)\G(1-u-v+\varepsilon_1) }{\G(\varepsilon_1) (-u-v+\varepsilon_1)} = -\frac{\G(\varepsilon_1) }{\G(\varepsilon_1)} = -1.
\end{eqnarray*}
This is the only residue in the complex $u$ plane that contributes, other residues  disappear in the limit of removing the regularization.    However, this residue before taking the limit was on the right hand side, it was a right residue, it contributes with the negative sign due to the clockwise orientation of the corresponding contour integral.   The contributions of other residues produced by  $\G(-u-v+\varepsilon_1)$ in the complex $u$ plane will vanish in this limit of the removing the regularization because $\G(\varepsilon_1)$ will not appear in their numerator,  there is  nothing to compensate  $\G(\varepsilon_1)$ in the denominator as it happened for the first residue at $u = \varepsilon_1 -v.$  Alternatively,  if we close the Mellin-Barnes integration contour to the left complex infinity for the variable $u,$  the situation will be the same. Only the first pole at $u =-v$ produced by  $\G(u+v)$ contributes in this case.  The contributions of other residues produced by
$\G(u+v)$  vanish  in the limit of removing the regularization  $\varepsilon_1 \rightarrow 0.$

 Thus, the five-fold Mellin-Barnes integral (\ref{five-2})  can be replaced with a four-fold Mellin-Barnes integral
\begin{eqnarray*}  
 \frac{1}{st}\int\limits_{-\delta_1 -i\infty}^{-\delta_1 + i\infty} ~dz_1\G(-z_1)\int\limits_{-\delta_2-i\infty}^{-\delta_2 + i\infty} ~dz_2 ~\G(-z_2)\int\limits_{1-\delta_1-\delta_2 -\delta_5-i\infty}^{1-\delta_1-\delta_2 -\delta_5  + i\infty} ~du~\G(1+z_1+z_2-u) \no\\
\int\limits_{\delta_1 -\delta_3 -i\infty}^{\delta_1 -\delta_3 + i\infty} ~dv~\G(-v-z_1)\int\limits_{1-\delta_1 -\delta_2 -\delta_4 -i\infty}^{ 1-\delta_1-\delta_2 -\delta_4 + i\infty} ~dw~\G(1+z_1+z_2-w)\G(u+w+v-z_2)\no\\ 
\le\frac{p_2^2 p_4^2}{st}\ri^{z_1}
\le\frac{p_3^2p_1^2}{st}\ri^{z_2}s^{w}t^u\le p_4^2\ri^{v} \le p_1^2\ri^{-v-w-u} \no\\
\int\limits_0^1~ d\alpha_4 ~ \alpha_4^{v+u-1}(1-\alpha_4)^{-1-u-v}
\int\limits_0^1 d\beta_3 (1- \beta_3)^{-v-w-u-1} \beta_3^{w-1} \int\limits_0^{1}d\gamma_2 ~(1-\gamma_2)^{-u-1}  \gamma_2^{-1-v-w}\no
\end{eqnarray*}  

\begin{eqnarray}  \label{four}
=\lim_{\varepsilon_1 \rightarrow 0}
\frac{1}{st}\int\limits_{-\delta_1 -i\infty}^{-\delta_1 + i\infty} ~dz_1\G(-z_1)\int\limits_{-\delta_2-i\infty}^{-\delta_2 + i\infty} ~dz_2 ~\G(-z_2)\int\limits_{1-\delta_1-\delta_2 -\delta_5-i\infty}^{1-\delta_1-\delta_2 -\delta_5  + i\infty} ~du~\G(1+z_1+z_2-u) \no\\
\int\limits_{\delta_1 -\delta_3 -i\infty}^{\delta_1 -\delta_3 +  i\infty} ~dv~\G(-v-z_1)\int\limits_{1-\delta_1 -\delta_2 -\delta_4 -i\infty}^{ 1-\delta_1-\delta_2 -\delta_4 + i\infty} ~dw~\G(1+z_1+z_2-w)\G(u+w+v-z_2)\no\\ 
\times\frac{\G(v+u)\G(-u-v+\varepsilon_1)}{\G(\varepsilon_1)}\le\frac{p_2^2 p_4^2}{st}\ri^{z_1}
\le\frac{p_3^2p_1^2}{st}\ri^{z_2}s^{w}t^u\le p_4^2\ri^{v} \le p_1^2\ri^{-v-w-u}  \no\\
\times\int\limits_0^1 d\beta_3 (1- \beta_3)^{-v-w-u-1} \beta_3^{w-1} \int\limits_0^{1}d\gamma_2 ~(1-\gamma_2)^{-u-1}  \gamma_2^{-1-v-w} \no \\
\no\\
=\frac{1}{st}\int\limits_{-\delta_1  -i\infty}^{-\delta_1 + i\infty} ~dz_1~\G(-z_1)  \int\limits_{-\delta_2-i\infty}^{-\delta_2   + i\infty} ~dz_2~ \G(-z_2)
\int\limits_{\delta_1-\delta_3 -i\infty}^{\delta_1 -\delta_3+i\infty   } ~dv 
\G(-v-z_1)
\G(1+z_1+z_2+v)\no\\
\int\limits_{1-\delta_1-\delta_2 -\delta_4 - i\infty}^{ 1-\delta_1-\delta_2 -\delta_4 + i\infty} ~dw~\G(1+z_1+z_2-w)\G(w-z_2)
\le\frac{p_2^2 p_4^2}{st}\ri^{z_1}\le\frac{p_3^2p_1^2}{st}\ri^{z_2}s^{w}t^{-v}  \no\\ 
\times\le p_4^2\ri^{v} \le p_1^2\ri^{-w}
\int\limits_0^1 d\beta_3 (1- \beta_3)^{-w-1} \beta_3^{w-1} \int\limits_0^{1}d\gamma_2 ~(1-\gamma_2)^{v-1} \gamma_2^{-1-w-v} 
\end{eqnarray}

The integral over $\beta_3$ is not convergent at the upper limit for  ${\rm Re~~} w$ defined by  Eq. (\ref{position}) but 
at the lower limit is convergent.    However, we know that the overall five-fold integral  (\ref{five-2})  is finite  because the original integral (\ref{Box-4D-alfa}) over the Feynman parameters  is finite.  This means that
\begin{eqnarray}  \label{4-beta}
\int\limits_0^1~ d\beta_3 ~ \beta_3^{w-1}(1-\beta_3)^{-1-w}
\end{eqnarray}
should be understood in a sense of  a distribution  over the new complex variable $w$ that is, the integral   (\ref{four})   can be concisely written as   
\begin{eqnarray}  \label{four-beta}
\int\limits_{1-\delta_1-\delta_2 -\delta_4 - i\infty}^{ 1-\delta_1-\delta_2 -\delta_4 + i\infty} ~dw~f_2(w)\int\limits_0^1 d\beta_3 (1- \beta_3)^{-w-1} \beta_3^{w-1} 
\end{eqnarray}
$f_2(w)$ here stands for the rest of the integrand in (\ref{four}).   The  four-fold integral   (\ref{four}) is not singular, 
it takes a finite value. 
This means that  (\ref{4-beta})   can be replaced in (\ref{four-beta}) with a  limit  which regularizes it, that is,
\begin{eqnarray*}  
\lim_{\varepsilon_2 \rightarrow 0} \int\limits_{1-\delta_1-\delta_2 -\delta_4 - i\infty}^{ 1-\delta_1-\delta_2 -\delta_4 + i\infty} ~dw~f_2(w)\int\limits_0^1 d\beta_3 (1- \beta_3)^{-w-1} \beta_3^{w-1 + \varepsilon_2}  \\
= \lim_{\varepsilon_2 \rightarrow 0} \int\limits_{1-\delta_1-\delta_2 -\delta_4 - i\infty}^{ 1-\delta_1-\delta_2 -\delta_4 + i\infty} ~dw~f_2(w)  \frac{\G(w)\G(-w+\varepsilon_2)}{\G(\varepsilon_2)}  = f_2(0), 
\end{eqnarray*}
where  we suppose that  we have started with   ${\rm Re~~}\varepsilon_2 > 2.$ 
Apparently, the pole  $w= \varepsilon_2 $  in the complex $w$ plane contributes because it is on the right hand side of the straight vertical line in the complex plane of the variable $w$ when
${\rm Re~~}\varepsilon_2 > 2,$ 
\begin{eqnarray*}
\res_{w= \varepsilon_2} \frac{\G(w)\G(-w+\varepsilon_2)}{\G(\varepsilon_2)} =
\res_{w= \varepsilon_2} \frac{\G(w)\G(1-w+\varepsilon_2) }{\G(\varepsilon_2) (-w+\varepsilon_2)} = -\frac{\G(\varepsilon_2) }{\G(\varepsilon_2)} = -1.
\end{eqnarray*}
This is the only residue in the complex $w$ plane that contributes, other residues  disappear then in the limit of removing the regularization. The contributions of other residues produced by  $\G(-w+\varepsilon_2)$ in the complex $w$ plane will vanish in this limit of the removing the regularization because $\G(\varepsilon_2)$ will not appear in their numerator,  there is nothing to    compensate  $\G(\varepsilon_2)$ in the denominator in comparison as it happened for the first residue at $w = \varepsilon_2 .$
However, this residue before taking the limit was on the right hand side of the straight line  of the integration contour in the complex plane $w,$ it was a "right" residue, it contributes with the negative sign due to the clockwise orientation of the corresponding contour integral.    Alternatively,  if we close the Mellin-Barnes integration contour to the left complex infinity for the variable $w,$  the situation will be the same. Only the first pole at $w =0$ produced by  $\G(w)$ contributes in this case.  The contributions of other residues produced by
$\G(w)$  vanish  in the limit of removing the regularization  $\varepsilon_2 \rightarrow 0.$

 Thus, the four-fold Mellin-Barnes integral (\ref{four}) integral can be replaced with three-fold Mellin-Barnes integral
\begin{eqnarray*} 
\frac{1}{st}\int\limits_{-\delta_1  -i\infty}^{-\delta_1 + i\infty} ~dz_1~\G(-z_1)  \int\limits_{-\delta_2-i\infty}^{-\delta_2  + i\infty} ~dz_2~ \G(-z_2)
\int\limits_{\delta_1-\delta_3 -i\infty}^{\delta_1-\delta_3 +i\infty   } ~dv 
\G(-v-z_1)\no\\
\G(1+z_1+z_2+v)\int\limits_{1-\delta_1-\delta_2 -\delta_4- i\infty}^{ 1-\delta_1-\delta_2 -\delta_4 + i\infty} ~dw~\G(1+z_1+z_2-w)\G(w-z_2) 
\le\frac{p_2^2 p_4^2}{st}\ri^{z_1}
\le\frac{p_3^2p_1^2}{st}\ri^{z_2} \no\\
s^{w}t^{-v}\le p_4^2\ri^{v} \le p_1^2\ri^{-w}
\int\limits_0^1 d\beta_3 (1- \beta_3)^{-w-1} \beta_3^{w-1} \int\limits_0^{1}d\gamma_2 ~(1-\gamma_2)^{v-1}\gamma_2^{-1-w-v}\no\\
  = \lim_{\varepsilon_2 \rightarrow 0}
\frac{1}{st}\int\limits_{-\delta_1 -i\infty}^{-\delta_1  + i\infty} ~dz_1~\G(-z_1)  \int\limits_{-\delta_2-i\infty}^{-\delta_2   + i\infty} ~dz_2~ \G(-z_2)
\int\limits_{\delta_1 -\delta_3 -i\infty}^{\delta_1 -\delta_3 +i\infty   } ~dv 
\G(-v-z_1)\no\\
\times \G(1+z_1+z_2+v)\int\limits_{1-\delta_1-\delta_2 -\delta_4 - i\infty}^{ 1-\delta_1-\delta_2 -\delta_4 + i\infty} ~dw~\G(1+z_1+z_2-w)\G(w-z_2)\no
\end{eqnarray*}
 
 \begin{eqnarray}  \label{three}
\le\frac{p_2^2 p_4^2}{st}\ri^{z_1}
\le\frac{p_3^2p_1^2}{st}\ri^{z_2}s^{w}t^{-v}\le p_4^2\ri^{v} \le p_1^2\ri^{-w}
 \frac{\G(w)\G(-w+\varepsilon_2)}{\G(\varepsilon_2)} \int\limits_0^{1}d\gamma_2 ~(1-\gamma_2)^{v-1}  \gamma_2^{-1-v-w} \no\\
 =  \frac{1}{st}\int\limits_{-\delta_1  -i\infty}^{-\delta_1 + i\infty} ~dz_1~\G(-z_1)  \int\limits_{-\delta_2-i\infty}^{-\delta_2  + i\infty} ~dz_2~ \G(-z_2)\no\\
\int\limits_{\delta_1-\delta_3-i\infty}^{\delta_1-\delta_3 +i\infty   } ~dv 
\G(-v-z_1)\G(1+z_1+z_2+v)\G(1+z_1+z_2)\G(-z_2)\no\\
\le\frac{p_2^2 p_4^2}{st}\ri^{z_1}\le\frac{p_3^2p_1^2}{st}\ri^{z_2}t^{-v}\le p_4^2\ri^{v}
\int\limits_0^{1}d\gamma_2 ~(1-\gamma_2)^{v-1}  \gamma_2^{-1-v} 
 \end{eqnarray}
The integral over $\gamma_2$ is not convergent at the upper limit  for ${\rm Re~~} v$ defined in Eq. (\ref{position}), but 
 at the lower limit is convergent  because the value of $v$ is situated on the straight line with the fixed real part 
 $\delta_1 -\delta_3.$
However, we know that the overall five-fold integral  (\ref{five})  is finite  because the original integral (\ref{Box-4D-alfa}) over the Feynman parameters  is finite.   This means that the integral
\begin{eqnarray} \label{4-gamma}
\int_0^1~ d\gamma_2 ~ \gamma_2^{v-1}(1-\gamma_2)^{-1-v}
\end{eqnarray}
should be understood in a sense of a distribution  over the new complex variables $v$ that is, the integral   (\ref{three})   can be concisely written as   
\begin{eqnarray}  \label{three-gamma}
 \int\limits_{\delta_1 -\delta_3 - i\infty}^{\delta_1 -\delta_3 +i\infty   }~dv f_3(v)\int\limits_0^1~ d\gamma_2 ~ \gamma_2^{v-1}(1-\gamma_2)^{-1-v}  
\end{eqnarray}
$f_3(v )$ here stands for the rest of the integrand in (\ref{three}).   The  three-fold integral   (\ref{three}) is not singular, 
it takes a finite value.   This means that  (\ref{4-gamma})   can be replaced in (\ref{three-gamma}) with a  limit  which regularizes 
 (\ref{4-gamma}),  that is,
\begin{eqnarray*}
\int\limits_{\delta_1 -\delta_3 -i\infty}^{\delta_1 -\delta_3 +i\infty   }~dv f_3(v)\int\limits_0^1~ d\gamma_2 ~ \gamma_2^{v-1}(1-\gamma_2)^{-1-v}  \\
= \lim_{\varepsilon_3 \rightarrow 0}\int\limits_{\delta_1 -\delta_3 -i\infty}^{\delta_1 -\delta_3+i\infty   } ~dv f_3(v)
\int\limits_0^1~ d\gamma_2 ~ \gamma_2^{v-1}(1-\gamma_2)^{-1-v +\varepsilon_3} \\
=  \lim_{\varepsilon_3 \rightarrow 0}\int\limits_{\delta_1 -\delta_3 -i\infty}^{\delta_1 -\delta_3 +i\infty   } ~dv f_3(v)
\frac{\G(v)\G(-v+\varepsilon_3)}{\G(\varepsilon_3)}  =  f_3(0).
\end{eqnarray*}
where  we suppose that  we have started with   ${\rm Re~~}\varepsilon_3 > 1.$ 
Apparently, the pole  $v= \varepsilon_3 $  in the complex $v$ plane contributes because it is on the right hand side of the straight vertical line in the complex plane of the variable $v$ when  ${\rm Re~~}\varepsilon_3 > 1,$ 
\begin{eqnarray*}
\res_{v= \varepsilon_3} \frac{\G(v)\G(-v+\varepsilon_3)}{\G(\varepsilon_3)} =
\res_{v= \varepsilon_3} \frac{\G(v)\G(1-v+\varepsilon_3) }{\G(\varepsilon_3) (-v+\varepsilon_3)}
= -\frac{\G(\varepsilon_3) }{\G(\varepsilon_3)} = -1.
\end{eqnarray*}
This is the only residue in the complex $v$ plane that contributes, other residues  disappear  in the limit of removing the regularization.  However, this  residue before taking the limit was on the right hand side of the straight line  of the integration contour in the complex plane $v,$ it was a "right" residue, it contributes with the negative sign due to the clockwise orientation of the corresponding contour integral.  Other residues in the complex plane of the variable $v$  contribute with  $\G(\varepsilon_3)$ in the denominator and  these contributions disappear when  we take the limit  $\varepsilon_3 \rightarrow 0.$  
 They are  produced by  $\G(-v+\varepsilon_3)$ in the complex $v$ plane and will vanish in this limit of the removing the regularization because $\G(\varepsilon_3)$ will not appear in their numerator,  there is nothing to    compensate  $\G(\varepsilon_3)$ in the denominator in comparison as it happened for the first residue at $v = \varepsilon_3 .$
Alternatively,  if we close the Mellin-Barnes integration contour to the left complex infinity for the variable $v,$  the situation will be the same. Only the first pole at $v =0$ produced by  $\G(v)$ contributes in this case.  The contributions of other residues produced by
$\G(v)$  vanish  in the limit of removing the regularization  $\varepsilon_3 \rightarrow 0.$

\section{Discussion} \label{sec-Discuss}

In quantum field theory, any Feynman diagram in any dimension can be expressed using Barnes integrals, whose integrands are ratios of Euler gamma functions.  In massless theories, the structure of these integrands is somewhat simplified because there are no dimensional parameters like scales or masses. However, even in these theories, the number of Mellin-Barnes  integration contours increases as the number of loops and external legs grows.  In this paper we considered a case when a number of contours is efficiently reduced from the five-fold  Mellin-Barnes to the two-fold Mellin-Barnes  integration.  This example  of a simple massless box diagram known already  for many years  helps us to elaborate a general strategy which may be used for many Feynman diagrams, with masses or massless.  Our observation highlights that certain integrals over Feynman parameters become singular after applying the Mellin-Barnes transformation to the original integrals over Feynman parameters. These singular integrals must be treated as distributions in the remaining Mellin-Barnes integrands. Analytical regularization is admitted  here because the overall integral remains convergent. Upon applying analytical regularization, these distributions transform into Euler beta functions. The contributions from these beta functions reduce to a finite number of residues after the analytical regularization is removed. These residues correspond to specific new complex variables, which are linear combinations of the initial complex integration variables from the Mellin-Barnes transformations. These linear combinations arise from the powers of the Feynman parameters in the singular integrals and can serve as new variables for the Mellin-Barnes transformations. The finite set of residues in the planes of these new variables is straightforward to evaluate, leaving us with Barnes integrals where the number of contour integrations is reduced. This reduction corresponds to the number of singular integrals over Feynman parameters after the Mellin-Barnes transformation applied to the denominator of the original integral over them.

The strategy for an arbitrary diagram is a quite straightforward generalization of the idea described in the previous paragraph. 
We take a Feynman diagram, apply the Feynman formula,   transform the denominator in the integrand over the Feynman parameters 
to the Mellin-Barnes integral, transform the integral over $n$-simplex to the integral over the unit volume (this is the unit $(n-1)$-cube where $n$ is the number of Feynman parameters)  and look which of the integrals over the Feynman parameters are singular. These singular integrals should be regularized by the analytical regularization. The regularized integrals result in the Euler beta functions which can be re-written as ratios of the Euler gamma functions. If in the denominator the gamma function of non-positive integer appears, the corresponding contour integral contributes with a finite number of residues. The true variables of these contour integrations coincide with powers of the  singular integrals over the Feynman parameters, which are certain linear combinations of the initial Mellin-Barnes integration variables.  The initial arguments of the Mellin-Barnes transformations can be re-grouped   according to these linear combinations after taking the contour integrals over these linear combinations off by evaluation a finite number of residues.   The rank of the special function of the re-grouped arguments  which appears after taking all the Mellin-Barnes integrations off, is reduced because  the number of the Mellin-Barnes integration contours is reduced.

The analytical regularization may be done only when the singular integrals  are parts of the  Mellin-Barnes integrands and the overall Mellin-Barnes integral  is convergent.   This is  the case usually  because it corresponds to a convergent overall integral over the Feynman
parameters.    Only in such a  case the singular integrals over the Feynman parameters  may be understood in a sense 
of distributions in the Mellin-Barnes integrands and may be regularized analytically.
For example, the left hand side  of (\ref{beta})  cannot be analytically regularized if the integral is divergent, this would  produce wrong results if done.  
At the same time, the inverse formula (\ref{Mellin-BarnesT-back-2}) is valid for any $\lambda$ even for  a negative integer.  This means that the integration domains of the left hand  sides of   (\ref{Mellin-BarnesT})   and   (\ref{beta})
should be modified for $\lambda$ negative    by   contour deformations  established in Refs. \cite{Kondrashuk:2019cwi,Alvarez:2019eaa,convergence}.  In turn, the modification of the left hand side of   (\ref{beta})
 would mean the modification of  the Feynman formula  (\ref{feynman-2})  to another formula  which will be valid after such a deformation of    (\ref{beta})  to arbitrary indices $\nu_i$ and dimension $d.$ As we have seen in Subsection \ref{subsec-two-fold-triangle}, the Feynman formula   (\ref{triangle-FP})
for the case of the triangle diagram  is valid for a wide domain of $\nu_i$ and $d$ determined by the requirement 
of convergence of  the integrals in  (\ref{triangle-FP}).  It is possible based on the technique developed in   \cite{Kondrashuk:2019cwi,Alvarez:2019eaa,convergence} to find how  the integral (\ref{triangle-FP}) should be modified in order to apply it for arbitrary indices and an arbitrary dimension.  However, for the four-point box diagram, the divergent integrals over Feynman parameters appear in the Mellin-Barnes integrand even for the  convergent overall integral   (\ref{Box-4D-alfa}) when all the indices  of the propagators are 1 and  the dimension $d =4$.
A modified Feynman formula would guarantee the convergence of these integrals in which the integration over the interval from 
0 to 1 would be replaced with the integration over  Pochhammer contour or over Hankel contour \cite{Kondrashuk:2019cwi,Alvarez:2019eaa,convergence}.  The analytical 
regularization in the Mellin-Barnes integrands is not necessary if such a modified Feynman formula is used.

The results obtained in this paper may be analyzed   from the perspective of  of knot theory.   Wick's theorem can be used to construct the Green functions  in quantum field theory \cite{hagen}.  Specifically, Wick’s theorem establishes a relationship between the Feynman path integral and knot theory  \cite{hagen}.  In particular, the relation between the  renormalization procedure and the knot theory was demonstrated  in  \cite{Kreimer}.  Indeed,  the Feynman diagrams can be classified in terms of the knot theory \cite{Kreimer}.    This implies that any correspondence between Feynman diagrams has a mirrored counterpart in knot theory, and vice versa. In particular, the loop reduction of ladder diagrams can be interpreted within knot theory \cite{Gonzalez:2016pgx,Kniehl:2013dma,Gonzalez:2012gu}.  Knot theory can also be described in terms of Jacobi diagrams \cite{Watanabe}.  For example, 
there are certain recurrent relations between the ladder diagrams with the neighbor  number of loops   \cite{Belokurov:1983km,Allendes:2012mr,Gonzalez:2016pgx,Kniehl:2013dma,Gonzalez:2012gu,Gonzalez:2012wk}.   These relations may be understood in terms of Jacobi diagrams within knot theory.  In general, the knot theory helps to evaluate periods of Feynman diagrams in terms of zeta functions 
 \cite{Schnetz,Brown:2024qhk}. Kontsevich’s result \cite{Kontsevich}  was employed to compute periods in    \cite{Schnetz,Brown:2024qhk}. 
 In \cite{Kontsevich} Kontsevich found a link between Feynman diagrams and Moyal star product of certain functions. The periods are used to calculate the renormalization constants   in quantum field theory.  However,   we do not know any bibliographic reference in which the knot theory would help to find an explicit result for a given Feynman diagram.

Quantum electrodynamics (QED), as one of the fundamental quantum field theories, may be analysed within the framework of knot theory \cite{SPIE}.  One-loop results in QED  after mapping to knot theory  can be represented in terms of knots  \cite{SPIE}.  
 At the beginning of QED,  Feynman evaluated the box diagram with electron propagators inside the loop  \cite{feyn-1,feyn-2,feyn-3,feyn-4,feyn-0}.   However, this case differs from the classical scenario as the electron possesses mass. 
 The massive off-shell scalar box diagram with four different masses is evaluated in  \cite{veltman,Hahn,oldenborg,Tarasov}.  
 In this paper, we consider the off-shell box diagram in the massless case to minimize the number of Mellin-Barnes integration contours. This reformulation allows for the minimal  representation of Mellin-Barnes complex integrals with minimal reliance on auxiliary tools, such as analytical regularization at intermediate steps.

 The evaluation of box Feynman diagrams exhibits an explicit connection to knot theory, particularly through the framework of Trotter integrals. The path integral in QED, and more generally in field theory, can be interpreted in terms of Trotter integrals \cite{feyn-1,feyn-2,feyn-3,feyn-4,feyn-0,Shulman}.   The approach based on the  Trotter integrals is widely applied  in  quantum  statistics, too \cite{dugave,dugave-2}. For example, it serves as a solution to Markov  process \cite{feynman-Kac,feynman-Kac-2}.
 By construction, Trotter integrals can also be formulated within the context of knot theory \cite{Trotter,Rozansky,Tyurina,kaufman,kaufman-2}.  
 Quantum mechanics  can be solved in terms of the Feynman path integral \cite{feyn-1,feyn-2,feyn-3,feyn-4,feyn-0,Shulman}. 
  If Feynman path integrals are treated as Trotter integrals, then our method can be extended to the computation of Trotter integrals in quantum mechanics. 
 Furthermore, Trotter integrals play a significant role in quantum mechanics, particularly in applications concerning quantum intentionality 
 \cite{Bulnes,Shaw}  and quantum computing  \cite{Bruntrup}.

\vspace{6pt}

\acknowledgments{M.D. is supported by the fellowship of ``Ayudante de investigacion de pregrado y postgrado 2021'' of VRIP UBB. He is grateful to Andrei Davydychev and Cristian Villavicencio for careful reading of his thesis and  for warm and constructive evaluation of his results during the thesis defence.
I.G. would like to thank CEFITEV-UV for partial support.
The work of I.K. was supported in part by Fondecyt (Chile) Grants Nos. 1040368, 1050512 and 1121030, by DIUBB (Chile) Grant Nos.  125009,  GI 153209/C  and GI 152606/VC. E.A.N.C. work was partially supported by DIDULS, Universidad de La Serena.
We thank all four referees of the present paper for their comments and suggestions which helped us to improve it.}




%

\appendixtitles{yes} 
\appendixstart
\appendix
\section[\appendixname~\thesection]{The original way from the Triangle to the Box}\label{App}
%
%
We reproduce here the trick of Usyukina and Davydychev \cite{Usyukina:1992jd} that they used  to relate the three-point and the four-point Green functions with only one purpose: to compare it with pure Mellin-Barnes approach which we proposed in the present paper.

The integral  (\ref{Box-4D-alfa})  can be developed
\begin{eqnarray*}
\int\limits_{0}^{1} \int\limits_{0}^{1}\int\limits_{0}^{1} \int\limits_{0}^{1}
d\alpha _{1}d\alpha _{2}d\alpha _{3}d\alpha _{4}
\frac{\delta \le \alpha _{1}+\alpha _{2}+\alpha _{3}+\alpha _{4}-1\ri }
{\left[ \alpha_1\le \alpha _{2}p_{1}^{2}+ \alpha _{3}s\ri  +
  \alpha_4  \le  \alpha_3 p_{3}^{2}  +  \alpha_2 t \ri
 +  \alpha_1\alpha_4 p_{4}^{2} + \alpha _{2}\alpha_{3}p_{2}^{2} \right] ^{2}} \\
= \int\limits_{0}^{1}d\alpha _{4}\int\limits_{0}^{1-\alpha _{4}}
d\alpha_{3}\int\limits_{0}^{1-\alpha _{4}-\alpha _{3}}d\alpha_{2}
\int\limits_{0}^{1-\alpha _{4}-\alpha _{3}-\alpha _{2}}d\alpha _{1}\;%
\frac{\delta \left( \alpha _{1}+\alpha _{2}+\alpha _{3}+\alpha _{4}-1\right)}
{\left[ \alpha _{1}X+\alpha _{4}Y  +  \alpha_1\alpha_4 p_{4}^{2} + \alpha _{2}\alpha_{3}p_{2}^{2}   \right]^2 },
\end{eqnarray*}
where we have introduced the notation for $X$  and $Y$
\begin{eqnarray*}
X = \alpha _{2}p_{1}^{2}+ \alpha _{3}s, ~~~ Y = \alpha_3 p_{3}^{2}  +  \alpha_2 t.
\end{eqnarray*}
After removing the integral over $\alpha_1$ and changing then the integration variables $\alpha_2$ and $\alpha_3$  as
\begin{eqnarray*}
\alpha_2 = (1-\alpha_4)\beta_2, ~~~~ \alpha_3 = (1-\alpha_4)\beta_3,
\end{eqnarray*}
we obtain

\begin{adjustwidth}{-\extralength}{0cm}
\centering 
\begin{eqnarray*}
\int\limits_{0}^{1}d\alpha _{4}\int\limits_{0}^{1-\alpha _{4}}
d\alpha_{3}\int\limits_{0}^{1-\alpha _{4}-\alpha _{3}}d\alpha_{2}
\frac{1}
{\left[ (1-\alpha_4-\alpha_3-\alpha_2) X+\alpha _{4}Y  +    (1-\alpha_4-\alpha_3-\alpha_2) \alpha_4 p_{4}^{2} + \alpha _{2}\alpha_{3}p_{2}^{2}   \right]^2 } \\
= \int\limits_{0}^{1}d\alpha _{4}\int\limits_{0}^{1}d\beta _{3}
\int\limits_{0}^{1-\beta _{3}}d\beta _{2}
\frac{\left( 1-\alpha_{4}\right) ^{2}}
{ \left[ \le 1-\alpha _{4}\ri^{2} \le (1-\beta _{3}-\beta _{2})\widetilde{X} + \beta _{2}\beta_{3}p_{2}^{2}
\ri   + \alpha _{4} \le 1-\alpha_{4}\ri \le  \widetilde{Y} +
\le 1-\beta _{3}-\beta _{2}\ri p_{4}^{2} \ri  \right]^{2}} \\
= \int\limits_{0}^{1}d\alpha _{4}\int\limits_{0}^{1}d\beta _{3}
\int\limits_{0}^{1-\beta _{3}}d\beta _{2}
\frac{1}
{ \left[ \le 1-\alpha _{4}\ri \le (1-\beta _{3}-\beta _{2})\widetilde{X} + \beta _{2}\beta_{3}p_{2}^{2}
\ri   + \alpha _{4}  \le  \widetilde{Y} + \le 1-\beta _{3}-\beta _{2}\ri p_{4}^{2} \ri  \right]^{2}}.
\end{eqnarray*}
\end{adjustwidth}
We have introduced another notation
\begin{eqnarray*}
\widetilde{X} = \beta _{2}p_{1}^{2}+ \beta _{3}s, ~~~ \widetilde{Y} = \beta_3 p_{3}^{2}  +  \beta_2 t
\end{eqnarray*}
Now we put again the $\delta$ function in order to create triple integration and perform the last integration over $\alpha_4$. Thus, we get

\begin{adjustwidth}{-\extralength}{0cm}
\centering 
\begin{eqnarray*}
\int\limits_{0}^{1}d\alpha_{4}\int\limits_{0}^{1}d\beta _{3}\int\limits_{0}^{1}d\beta_{2}
\int\limits_{0}^{1}d\beta_1\frac{\delta(\beta_1+\beta_2+\beta_3-1)}
{ \left[ \le 1-\alpha_{4}\ri \le \beta_1\widetilde{X} + \beta _{2}\beta_{3}p_{2}^{2} \ri
+ \alpha _{4}  \le  \widetilde{Y} + \beta_1 p_{4}^{2} \ri  \right]^{2}} \\
=\int\limits_{0}^{1} \frac{ d\beta_{3}d\beta _{2}d\beta _{1} \delta \le \beta_{1}+
\beta_{2}+\beta _{3}-1\ri }{\beta_{1}\widetilde{X} + \beta_{2}\beta_3 p_{2}^{2}  -\widetilde{Y} -\beta _{1}p_{4}^{2}}
\left[ \frac{1}{\widetilde{Y} +\beta_{1} p_{4}^{2}}
-\frac{1}{\beta_{1}\widetilde{X} + \beta_{2}\beta_3 p_{2}^{2}}\right] \\
=\int\limits_{0}^{1} \frac{\delta \le \beta_{1}+
\beta_{2}+\beta _{3}-1\ri d\beta_{3}d\beta _{2}d\beta _{1}}{\le \beta_{1}\widetilde{X} + \beta_{2}\beta_3 p_{2}^{2}\ri  \le \widetilde{Y} + \beta _{1}p_{4}^{2} \ri }
= \int\limits_{0}^{1} \frac{\delta \le \beta_{1}+
\beta_{2}+\beta _{3}-1\ri d\beta_{3}d\beta _{2}d\beta _{1}}{\le \beta_{1} \beta _{2}p_{1}^{2}+  \beta_1\beta _{3}s  + \beta_{2}\beta_3 p_{2}^{2}\ri
\le \beta_3 p_{3}^{2}  +  \beta_2 t + \beta _{1}p_{4}^{2} \ri }
\end{eqnarray*}
\end{adjustwidth}

This integral can be transformed to the two-fold Mellin-Barnes integral by representing the second factor in the denominator in terms of the two-fold Mellin-Barnes integral
and by using the uniqueness formula for the integrals over the simplex, 
\begin{eqnarray} \label{UD1}
\int\limits_0^1\int\limits_0^1\int\limits_0^1 \frac{\delta\le\sum\limits_{i=1}^3 \beta_i -1\ri\prod\limits_{i=1}^3d\beta_i}
{(\beta_1\beta_2p_1^2 + \beta_2\beta_3 p_2^2  + \beta_1\beta_3 s) ( \beta_3 p_3^2 + \beta_1 p_4^2 + \beta_2 t )} \\
= \frac{1}{st}\oint\limits_C dz_1~dz_2 \le\frac{p^2_1p^2_3}{st}\ri^{z_1}  \le\frac{p^2_2p^2_4}{st}\ri^{z_2} \G^2 \le -z_1 \ri \G^2 \le -z_2 \ri \G^2 \le 1 + z_1 + z_2 \ri  \no
\end{eqnarray}

\begin{adjustwidth}{-\extralength}{0cm}

\reftitle{References}

\PublishersNote{}
\end{adjustwidth}
\end{document}